\documentclass[12pt]{iopart}
\usepackage[square,numbers,sort&compress]{natbib}
\usepackage{subcaption}
\usepackage{graphicx}
\usepackage{hyperref}
\usepackage{amssymb}

\usepackage{braket}
\usepackage{booktabs}
\usepackage{xcolor}
\newif\ifrevision
\newcommand{\rev}[1]{\ifrevision\textcolor{red}{#1}\else#1\fi}
\newenvironment{revision}
{\ifrevision\color{red}\fi}
{}

\makeatletter
\newcommand{\appsection}[2]{%
  \refstepcounter{section}%
  \setcounter{subsection}{0}%
  \section*{Appendix~\Alph{section}. #1}%
  \label{#2}%
}
\newcommand{\appsubsection}[2]{%
  \refstepcounter{subsection}%
  \subsection*{\Alph{section}.\arabic{subsection}\ #1}%
  \label{#2}%
}
\makeatother

\begin{document}

\title{Nonstabilizerness Estimation using \\Graph Neural Networks}

\author{Vincenzo Lipardi$^1$*, Domenica Dibenedetto$^1$, \\Georgios Stamoulis$^1$, Evert van Nieuwenburg$^2$ and \\Mark H.M. Winands$^1$}

\address{$^1$Department of Advanced Computing Sciences, Maastricht University, 6229 EN Maastricht, The Netherlands}
\address{$^2$ Leiden Institute of Advanced Computer Science, Leiden University, 2333 CA Leiden, The Netherlands}
\address{}

\eads{\mailto{vincenzo.lipardi@maastrichtuniversity.nl}}

\begin{abstract}
This article proposes a Graph Neural Network (GNN) approach to estimate nonstabilizerness in quantum circuits, measured by the stabilizer Rényi entropy (SRE). Nonstabilizerness is a fundamental resource for quantum advantage, and efficient SRE estimations are highly beneficial in practical applications. 
We address the nonstabilizerness estimation problem through three supervised learning formulations starting from easier classification tasks to the more challenging regression task. 
Experimental results show that the proposed GNN manages to capture meaningful features from the graph-based circuit representation, resulting in robust generalization performances achieved across diverse scenarios.
In classification tasks, the GNN is trained on product states and generalizes on circuits evolved under Clifford operations, entangled states, and circuits with higher number of qubits. In the regression task, the GNN significantly improves the SRE estimation on out-of-distribution circuits with higher number of qubits and gate counts compared to previous work, for both unstructured random quantum circuits and structured circuits derived from the transverse-field Ising model. Moreover, the graph representation of quantum circuits naturally integrates hardware-specific information. Simulations on noisy quantum hardware highlight the potential of the proposed GNN to predict the SRE measured on quantum devices.
\end{abstract}

%

\section{Introduction}
Quantifying the computational resources required to achieve a quantum advantage is a main problem in quantum information theory. Nonstabilizerness, also known as \textit{magic}~\cite{bravyi2005universal, campbell2017roads}, emerges as a key resource for quantum advantage as it measures the deviation of a quantum state from the polynomial-time simulability in the framework of the Gottesman-Knill Theorem~\cite{gottesman1998heisenberg, aaronson2004improved}. 
This theorem states that any quantum computation that consists exclusively of Clifford operations can be perfectly simulated in polynomial time by a classical machine. However, preparing a generic quantum state in the full Hilbert space requires resources beyond the Clifford group, since a universal gate set cannot be constructed using Clifford gates alone. The amount of these non-Clifford resources is referred to as nonstabilizerness.  

To rigorously capture the effect of non-Clifford operations in quantum computations, several measures of nonstabilizerness have been proposed. Examples include negative quasi-probability~\cite{veitch2012negative}, robustness of magic~\cite{howard2017application,heinrich2019robustness}, stabilizer nullity~\cite{beverland2020lower}, Bell magic~\cite{haug2023scalable}, and stabilizer R\'enyi entropies~\cite{leone2022stabilizer,leone2024stabilizer}.
Among these measures, we focus on the Stabilizer R\'enyi Entropy (SRE)~\cite{leone2022stabilizer} for two main reasons. First, SRE offers favorable computational properties~\cite{leone2022stabilizer,leone2024stabilizer,ahmadi2024quantifying}, especially when compared to alternative measures that require costly optimization procedures~\cite{veitch2012negative,howard2017application,heinrich2019robustness}. Second, SRE is particularly suitable for experiments on real quantum devices~\cite{oliviero2022measuring,niroula2024phase}. 
However, in the general case, the computational time required to compute the SRE of a quantum state scales exponentially in the number of qubits~\cite{leone2022stabilizer}.
Tensor Network (TN) methods are currently the most widely adopted approach to compute SRE for quantum states~\cite{haug2023quantifying,tarabunga2024nonstabilizerness,tarabunga2023many,lami2023nonstabilizerness,lami2024unveiling}. Although TN methods offer a theoretical framework with polynomial scaling, they are practically limited to low bond dimension, which restricts the class of quantum states that can be efficiently represented. Recently, Neural Quantum States~\cite{sinibaldi2025non} have been proposed as a promising alternative to overcome these limitations by flexibly approximating quantum states with high entanglement.
Motivated by previous research in machine learning applications to quantum technology~\cite{carleo2019machine,dunjko2018machine,krenn2023artificial}, learning-based approaches have recently been proposed for trading off online computational cost with approximate nonstabilizerness estimation. On one hand, in the restricted case of the stabilizer state classification, where the task is to distinguish between stabilizer and magic states~\cite{mello2025retrieving}, Convolutional Neural Networks achieve predictions that are robust under the Clifford group. On the other hand, classical machine learning approaches trained on tabular data achieve accurate estimation of SRE, but with limited generalization performances~\cite{lipardi2025study}.

In this article, we formulate the nonstabilizerness estimation as a graph-level prediction task and propose a Graph Neural Network (GNN)~\cite{scarselli2008graph} approach to bridge the gap between machine learning models that achieve robust performance on the restricted classification task and models that are capable of estimating SRE on the general regression task but with limited generalization performances. 

The three main contributions of the article are as follows.
\begin{enumerate}
    \item  We propose a GNN approach for the nonstabilizerness estimation problem, that exhibits higher generalization capabilities to larger quantum circuits, in terms of number of qubits and gate counts, and to unseen entanglement, compared to previous works~\cite{mello2025retrieving,lipardi2025study}. The experimental results highlight the positive impact of representing quantum circuits as graphs, which enables the model to effectively capture their intrinsic structural properties. 
    \item We have generated and released a comprehensive dataset of quantum circuits tailored to address the nonstabilizerness estimation problem in the supervised learning setting~\cite{lipardi2025dataset}. The dataset contains a total of $1686000$ quantum circuits designed to span a broad range of number of qubits, entanglement and SRE values. The dataset is structured to address the problem in three different formulations of increasing difficulty, including both classification and regression tasks. We expect this dataset to foster progress in the development and benchmarking of machine learning methods for nonstabilizerness estimation. 
    \item We show the proposed GNN model is well-suited to predict the SRE measured on real quantum devices, as the graph-based representation can also encode the hardware specifics.
\end{enumerate}
We note that the proposed approach trades off the quality of SRE predictions for a significantly reduced prediction time~\cite{lipardi2025study}. In fact, after a one-time training procedure, the GNN can approximate SRE for a quantum circuit almost in real time. As such, it should be regarded as a complementary approach to TN methods rather than a competing one. This approach is relevant in applications where an exact computation of $M_2$ can be traded off for a faster approximation. Examples include exploring the relationship between favorable solutions in variational quantum algorithms and their corresponding SRE values~\cite{spriggs2025quantum}, as well as enabling quantum architecture search methods to guide the circuit design towards quantum circuits with high SRE values.

This article is structured as follows. Section~\ref{background} introduces the nonstabilizerness estimation problem. Section~\ref{related_work} discusses related work in the field of machine learning.
Section~\ref{data_generation} describes the data generation procedure and details the rationale behind its design.
Section~\ref{methods} describes the architecture and implementation details of the proposed GNN model.
Section~\ref{results} presents the experimental setup and discusses the results. 
Section~\ref{conclusions} summarizes the main findings of this research study, highlighting the strengths and limitations of the proposed model and outlining promising directions for future research.

\section{Background} \label{background}
This section introduces the stabilizer formalism in Section~\ref{stabilizer_states}, defines the SRE and describes the main challenges in its estimation in Section~\ref{sre}, and concludes with the problem formulations of the nonstabilizerness estimation problem presented in Section~\ref{problem_formulation}.

\subsection{Stabilizer States} \label{stabilizer_states}
Let us consider an $n$-qubit quantum state and let $U(n)$ be the group of unitary operators on the Hilbert space $\mathcal{H}= (\mathbb{C}^2)^{\otimes n}$. We define the Pauli group $\mathcal{\widetilde{P}}_n$ acting on $\mathcal{H}$ as the set of all possible Pauli strings $\mathcal{P}_n$ taken with multiplicative factors $\pm 1$ and $\pm i$, and the Clifford group $\mathcal{C}(n) \subset U(n)$ as the normalizer of the Pauli group \cite{nielsen2010quantum}:
\begin{equation}
    \mathcal{\widetilde{P}}_n = \{\{\pm 1, \pm i\}\times \mathcal{P}_n \} \; ,\label{pauli_group}
\end{equation}
\begin{equation}
    \mathcal{C}(n)=
\{ C \in U(n) \;|\; \forall P \in \mathcal{\widetilde{P}}_n,\;
C P C^\dagger = P' \in \mathcal{\widetilde{P}}_n \}\;. \label{clifford_group}
\end{equation}
Let us recall that a Pauli string is a tensor product of Pauli matrices, each acting on a different qubit of an $n$-qubit system: $\mathcal{P}_n=\{\sigma_0,\sigma_1,\sigma_2,\sigma_3\}^{\otimes n}$. $\mathcal{\widetilde{P}}_n$ and $\mathcal{C}(n)$ are groups under the operation of matrix multiplication.
The set of gates consisting of the \texttt{CNOT} gate, the Hadamard gate \texttt{H}, and the phase gate \texttt{S} is a generator of the Clifford group $\mathcal{C}(n)$. 
Let $\ket{b}$ be any computational basis vector of $\mathcal{H}$. We can define the set of stabilizer states as the full Clifford orbit of $\ket{b}$~\cite{nielsen2010quantum}:
\begin{equation}
\mathrm{STAB} =\{ C \ket{b}  | \;C \in \mathcal{C}(n) \}. \label{stab}
\end{equation}

As proved by the Gottesman-Knill Theorem~\cite{gottesman1998heisenberg}, stabilizer states can be perfectly simulated by a classical computer in polynomial time. As a consequence, quantum computations restricted to stabilizer states and Clifford gates do not provide any computational advantage over the classical counterpart. 
Because the set of stabilizer states is closed under Clifford operations, and a universal quantum gate set cannot be constructed from Clifford gates alone, the preparation of a generic state in the Hilbert space requires resources beyond the Clifford group. The amount of such non-Clifford resources is referred to as the \textit{nonstabilizerness} of the state.

Since it is possible to prepare fully-entangled quantum states that can be efficiently simulated on a classical computer~\cite{nielsen2010quantum,aaronson2004improved}, entanglement is not the only resource to consider in the quest for quantum advantage. Nonstabilizerness, also known as \textit{magic}, plays a fundamental role in assessing quantum advantage~\cite{gottesman1998heisenberg, bravyi2005universal,campbell2017roads}, and, as such, it is important to define a measure for it. A proper measure of nonstabilizerness has to be a nonstabilizer (magic) monotone, as rigorously studied in magic-state resource theory~\cite{chitambar2019quantum, heimendahl2022axiomatic, liu2022many}.

   
The first measures of nonstabilizerness that have been proposed involve optimization procedures~\cite{veitch2012negative,howard2017application,heinrich2019robustness}, which come with high computational cost.
Then, more tractable measures have been introduced, such as stabilizer nullity~\cite{beverland2020lower}, Bell magic~\cite{haug2023scalable}, and stabilizer R\'enyi entropies~\cite{leone2022stabilizer}. 
In this article, we focus on the stabilizer R\'enyi entropy (SRE)~\cite{leone2022stabilizer} due to its favorable computational properties~\cite{haug2023quantifying,tarabunga2023many,lami2023nonstabilizerness,lami2024unveiling,ahmadi2024quantifying}, especially when compared to alternative measures that require costly optimization procedures~\cite{veitch2012negative,howard2017application,heinrich2019robustness}, and its practical suitability for experimental measurements on quantum hardware~\cite{oliviero2022measuring, niroula2024phase}.

\subsection{Stabilizer Rényi Entropy} \label{sre}
For a pure $n$-qubit quantum state $\rho$, the \textit{Stabilizer R\'enyi Entropy} (SRE)~\cite{leone2022stabilizer} of order $\alpha$ (also called R\'enyi index) is defined as:
\begin{equation}
    M_{\alpha} (\rho)= \frac{1}{1-\alpha} \ln \sum_{P\in \mathcal{P}_n}\Xi^{\alpha}_P(\rho) -\ln(2^n) \label{sre_formula}
\end{equation}
\noindent where $\Xi_P(\rho)= \frac{1}{2^n}Tr(\rho P)^2$ and $\mathcal{P}_n$ denotes the set of $n$-qubit Pauli strings. $M_{\alpha}$ is a nonstabilizerness monotone for R\'enyi index $\alpha\geq 2$~\cite{leone2024stabilizer, haug2023stabilizer}. In this article, we fix the index to the commonly used $\alpha=2$~\cite{leone2024stabilizer}, and we will refer to SRE simply as $M_2$. We note that $M_2$ is additive~\cite{leone2022stabilizer}, which means that for any quantum state $\ket{\psi_1}$ and $\ket{\psi_2}$: $M_2(\ket{\psi_1} \otimes \ket{\psi_2}) = M_2(\ket{\psi_1}) + M_2(\ket{\psi_2})$. 

Methods based on Matrix Product States~\cite{haug2023quantifying,lami2023nonstabilizerness,lami2024unveiling,tarabunga2024nonstabilizerness}, and more in general Tensor Networks, offer a framework to compute $M_2$ efficiently in the number of qubits but in running time that scales as a polynomial of degree four in the bond dimension~\cite{haug2023quantifying}. 
These methods represent a powerful theoretical framework, but with practical limitations to low bond dimensions~\cite{tarabunga2023many}. To overcome those limitations, recent advances have been made towards efficient $M_2$ computations using Neural Quantum States~\cite{sinibaldi2025non}. However, in the general case, the computational cost to compute $M_2$ of a quantum state using the Equation~\ref{sre_formula} grows exponentially with respect to the number of qubits $n$~\cite{leone2022stabilizer}, as it requires estimating $4^n$ expectation values, corresponding to all possible combinations of Pauli strings. 

The SRE generalization to mixed states $\tilde{M_2}$ is defined by rescaling $M_2$ with the logarithm of the purity of the mixed quantum state $\rho$~\cite{leone2022stabilizer, leone2024stabilizer, oliviero2022measuring}:
\begin{equation}
    \tilde{M_2}(\rho)= M_2(\rho)-\ln Tr(\rho^2) \label{sre_mixedstates}.
\end{equation}
 We note that $\tilde{M_2}$ is zero on the stabilizer mixed state defined as the equally-weighted mixed state over all the $n$-qubits stabilizer states\rev{, but it might be greater than zero for some mixtures of stabilizer states.}
\rev{Since $\tilde{M}$ is not a proper nonstabilizer monotone for mixed states, we also consider the witness of magic $\mathcal{W}_2$~\cite{haug2026efficient}, which satisfy the relation:
\begin{equation}
    \mathcal{W}_2(\rho)= \tilde{M}_2(\rho)+ 2 \ln Tr(\rho^2). \label{witness}
\end{equation}
A positive value of $\mathcal{W}_2({\rho})$ certifies that the state $\rho$ lies outside the stabilizer polytope, e.g. it is a nonstabilizer state. However, $\mathcal{W}_2({\rho})\leq 0$ does not guarantee that $\rho$ is a stabilizer state~\cite{haug2026efficient}.
}

We refer the interested reader to the existing literature for further details on stabilizer monotones for mixed states~\cite{leone2022stabilizer, haug2023stabilizer, leone2024stabilizer, yashin2025characterization, haug2026efficient, sonya2025nonstabilizerness}.

\subsection{Problem Formulation} \label{problem_formulation}
In this section, we define three formulations of the nonstabilizerness estimation problem on quantum circuits, with the goal of providing tasks of increasing difficulty that can serve as benchmarks for our study and future research.
The first is a classification task aimed at distinguishing quantum circuits that prepare stabilizer states from those that prepare magic states. This formulation relies uniquely on the definition of stabilizer states and it is therefore independent from all measures of nonstabilizerness, including $M_2$. We refer to this task as the stabilizer state classification. 
The second is a classification task based on SRE~\cite{leone2022stabilizer}, where the goal is to distinguish between quantum circuits with low and high $M_2$ values. The difficulty of this task is parameterized by the $M_2$ threshold, which determines the boundary between low-$M_2$ and high-$M_2$ quantum circuits. We refer to this task as the SRE-based classification. 
Finally, the third formulation is a regression task, where the goal is to estimate the degree of nonstabilizerness of a quantum circuit by predicting its positive real-valued $M_2$. We refer to this task as the SRE estimation. 

We emphasize that our formulation targets nonstabilizerness estimation at the circuit level, rather than at the level of quantum states. As a consequence, it does not require any form of quantum state tomography and relies uniquely on the circuit-based quantum circuit representation. 
On the one hand, the circuit-based formulation offers significant advantages for applications that are inherently defined within the circuit-model paradigm of quantum computing. An example are Variational Quantum Algorithms~\cite{cerezo2021variational}, whose link to $M_2$ has been recently discussed in~\cite{spriggs2025quantum}. On the other hand, the circuit-based formulation is not directly applicable to scenarios where the $M_2$ must be evaluated on unknown quantum states, for which a corresponding circuit is generally unavailable. 

The stabilizer state classification, particularly relevant in the context discussed in~\cite{mello2025retrieving}, becomes efficiently solvable in its matrix or circuit formulation~\cite{de2025fast}.  
Differently from the stabilizer state classification, the SRE-based classification and the SRE estimation both involve evaluating the contribution of each gate and its parameters to the overall circuit $M_2$, depending on their positions, which is generally hard. 
The SRE-based classification is not only computationally difficult but also physically relevant. In experiments on real quantum hardware, quantum noise can affect the measured $M_2$ values depending on both the circuit structure and the hardware specifics. By setting an appropriate threshold close to zero, the SRE-based classification can be used to identify stabilizer states in the presence of quantum noise. 
Lastly, the SRE estimation is the more general and challenging formulation of the nonstabilizerness estimation problem. 


\section{Related Work}\label{related_work}
In the last decade, different review articles have emphasized the growing influence of machine learning in the physical sciences, see for example~\cite{carleo2019machine} and the reference therein, particularly within the rapidly evolving field of quantum technologies~\cite{dunjko2018machine,krenn2023artificial}.

In quantum computing, particular efforts have been made in tackling problems such as fidelity estimation~\cite{zhang2021direct,wang2022torchquantum, saravanan2022data, vadali2024quantum}, quantum compiling~\cite{acampora2021deep, zlokapa2020deep, zhang2020topological, fan2022optimizing,ruiz2025quantum}, ground-state energy estimation in many-body quantum systems~\cite{lewis2024improved}, quantum error mitigation~\cite{liao2024machine}, quantum circuit expressibility estimation~\cite{aktar2024graph}. 
Following this research line, supervised learning approaches have been recently proposed to estimate the nonstabilizerness of quantum states~\cite{mello2025retrieving, lipardi2025study}. The first is a Convolutional Neural Network (CNN) approach proposed to distinguish between stabilizer and magic states~\cite{mello2025retrieving}. In this work, the authors construct a balanced dataset of stabilizer and magic states restricted to product states of fixed size, with $18$ qubits. Each state is represented by a $\mathcal{N}\times n$ binary matrix, where each row corresponds to a single-shot measurement outcome in the computational basis. This matrix is then converted into an image by assigning white pixels to zeros and black pixels to ones, enabling the use of CNN architectures.
The CNN model achieves promising results, including predictions that are robust under the Clifford group, which means good generalization capabilities on entangled states. However, the CNN approach is limited to the stabilizer state classification problem and to fixed size of quantum states, because the number of qubits affects the size of the images. The second approach extended the nonstabilizerness estimation problem to a regression task for the supervised machine learning setting~\cite{lipardi2025study}. For this purpose, they use the $M_2$ as a measure of nonstabilizerness for quantum circuits. In the general case, computing $M_2$ of quantum states scales exponentially with the qubit number. The authors tackle the SRE estimation problem with different machine learning models, including Random Forest Regressor and Support Vector Regressor, providing a study on various tabular classical representations of quantum circuits, including circuit-level feature and expectation values of $1$ and $2$-local observables calculated using classical shadows~\cite{huang2020predicting}. The Support Vector Regressor (SVR) trained on feature vectors encoding the gate counts achieve the best performance. Altough this approach showed promising experimental results to trade off online computational cost for approximate $M_2$ with machine learning, the models exhibit limited generalization capabilities to unseen larger circuits, in terms of both number of qubits and total gate count.

In this article, we advance beyond both the CNN and SVR approaches by proposing a Graph Neural Network (GNN) approach for nonstabilizerness estimation. Our method naturally accommodates both classification and regression formulations of the problem, while offering improved generalization capabilities compared to previous works~\cite{mello2025retrieving, lipardi2025study}. 
It is important to note that, in general, more complex models such as GNNs do not necessarily guarantee better performance. For example, in a recent study on supervised machine learning approaches to quantum error mitigation, Random Forests were shown to outperform other models, including GNNs~\cite{liao2024machine}. The choice of data representations and model architectures is often task-dependent, and the design of the appropriate combination of features and models facilitate the learning of relevant properties and ensure robust generalization performance.

\section{Data Generation}\label{data_generation}
We generate a dataset divided into three main subsets, each corresponding to one of the problem formulations introduced in Section~\ref{problem_formulation}.
In particular, Section~\ref{data_classification} describes the procedure for generating data for the stabilizer state classification problem, Section~\ref{data_classification_sre} for the SRE-based classification, and Section~\ref{data_regression} for the SRE estimation problem. Additionally, Appendix~\ref{data_generation_appendix} presents supplementary analysis of the datasets obtained.

\subsection{Data for Stabilizer State Classification}\label{data_classification}
The primary goal of this section is to introduce a well-defined and publicly available dataset for stabilizer state classification, which can serve as a benchmark for evaluating and comparing machine learning models under different generalization scenarios. Unlike previous work~\cite{mello2025retrieving}, where stabilizer and magic states exhibit clear separability in terms of their $M_2$ values, our proposed procedure ensures the generation of magic states spanning a broad range of $M_2$ values, including low-$M_2$ states that are more challenging to classify. It is important to note that, while the stabilizer state classification is an interesting task in measurement-based representations, it becomes a trivial problem in our circuit-based formulation~\cite{de2025fast}. 

The dataset for stabilizer state classification is constructed to formulate a circuit-level classification problem, where quantum circuits are labeled according to the quantum states they prepare, specifically distinguishing between stabilizer and magic states. Stabilizer states are assigned the label 0, while magic states are assigned the label 1.
All circuits prepare $n$-qubit states, initialized as $\ket{0}^{\otimes n}$, where the number of qubits $n$ spans the range from $2$ to $25$. The dataset consists of three main subsets, differing by the circuit composition and degree of entanglement:
\begin{itemize}
    \item Product States (PS): quantum circuits composed only of single-qubit rotations gates: \texttt{RX}, \texttt{RY}, and \texttt{RZ}.
    \item Clifford-evolved States (CS): for each circuit in PS we generate $25$ new circuits by evolving it under Clifford gates, uniformly sampled from the Clifford generator (\texttt{H}, \texttt{S}, \texttt{CNOT}). The Clifford gates are appended sequentially at the end of the product-state circuit. Each of the new circuits has a progressively higher number of new Clifford gates, referred as Clifford depth, from $1$ to $25$. 
    \item Entangled States (ES): for each circuit in PS we generate one new circuit injected with a number of \texttt{CNOT} gates uniformly sampled between $1$ and $20$ and randomly placed throughout the circuit. The placement of each \texttt{CNOT} gate involves the random selection of two qubits to act as control and target, respectively. To ensure a non-trivial entangling structure, qubits that have not yet been acted upon by any gate are excluded from being selected as control qubits.
\end{itemize}

Note that, since PS is balanced between the two labels and because nonstabilizerness is invariant under Clifford, CS and ES are also balanced. Furthermore, by construction, the circuits in CS preserve part of the structure of the circuits in PS, while the circuits in ES may result in strongly different quantum circuit structures due to the random placement of \texttt{CNOT} gates rather than their sequential appending. The computation of $M_2$ for circuits in PS can be easily performed by leveraging its additivity and compute $M_2$ as the sum of the $M_2$ computed per each qubit. Although CS inherits the $M_2$ values from the PS subset, the ES subset does not, since the contribution of each gate to the overall $M_2$ can change after the random \texttt{CNOT} injections, in the general case. 
We refer the reader to the Section~\ref{data_classification_appendix} for a detailed analysis of the data generated for the stabilizer state classification problem.

\subsection{Data for SRE-based Classification}\label{data_classification_sre}
The datasets used for the SRE-based classification task derive from those introduced in the previous section for the stabilizer state classification.

All stabilizer states are removed from the original dataset, leaving only the quantum circuits labeled as magic states (with label one). Subsequently, the median of the $M_2$ values of the remaining circuits is computed and taken as the threshold $M_2$ to distinguish between circuits with low and high $M_2$.
This procedure makes the classification task particularly challenging, as the median corresponds to a region densely populated by quantum circuits with similar $M_2$ values.
We refer the reader to the Section~\ref{data_classification_sre_appendix} for a detailed analysis of the data generated for the SRE-based classification problem.

\subsection{Data for Stabilizer Rényi Entropy Estimation} \label{data_regression}
The procedure used to generate the dataset for the regression formulation of the SRE estimation problem follows the approach described in~\cite{lipardi2025study}, and reported in the following. In addition, we extend the dataset to noisy quantum simulations to investigate the effect of quantum noise on the performance of our model. The dataset comprises two classes of quantum circuits, both defined for systems with a number of qubits $n$ ranging from $2$ to $6$. The first class includes \(250000\) random quantum circuits (unstructured), \(50000\) for each value of \( n \), and is hereafter referred to as the RQC dataset. The second class includes \(25000\) quantum circuits based on the Trotterized dynamics of the one-dimensional transverse-field Ising model (structured), \(5000\) for each value of \( n \), hereafter referred to as the TIM dataset. All circuits are labeled with their SRE values, computed from Equation~\ref{sre_formula} \rev{and Equation~\ref{sre_mixedstates}}. Each circuit is associated with two labels: the first corresponds to the $M_2$ obtained from noiseless simulations using a statevector simulator, and the second to the \rev{$\tilde{M_2}$} obtained from noisy simulations emulating the \texttt{FakeOslo} backend of IBM Quantum. This dual labeling enables a systematic analysis of the impact of quantum noise on the GNN model performance.

These datasets are designed to be representative of typical circuits used in applications for NISQ devices~\cite{preskill2018quantum}, covering a wide range of $M_2$ values. The variety of quantum circuit structures included in the dataset is crucial for studying the generality and robustness of our approach. On the one hand, random quantum circuits provide a general and unstructured benchmark, as they are not linked to any specific problem domain. On the other hand, the transverse-field Ising model serves as a well-established benchmark in quantum machine learning, owing to its fundamental role in condensed matter physics, statistical mechanics, and quantum information. Moreover, the behavior of the SRE in the transverse-field Ising spin chain has been extensively investigated~\cite{oliviero2022magic}.
We refer the reader to the Section~\ref{data_regression_appendix} for a detailed analysis of the data generated for the SRE estimation problem.

\section{Graph Neural Networks for Nonstabilizerness Estimation} \label{methods}
\begin{figure}[!hb]
    \centering
    \includegraphics[width=0.99\linewidth]{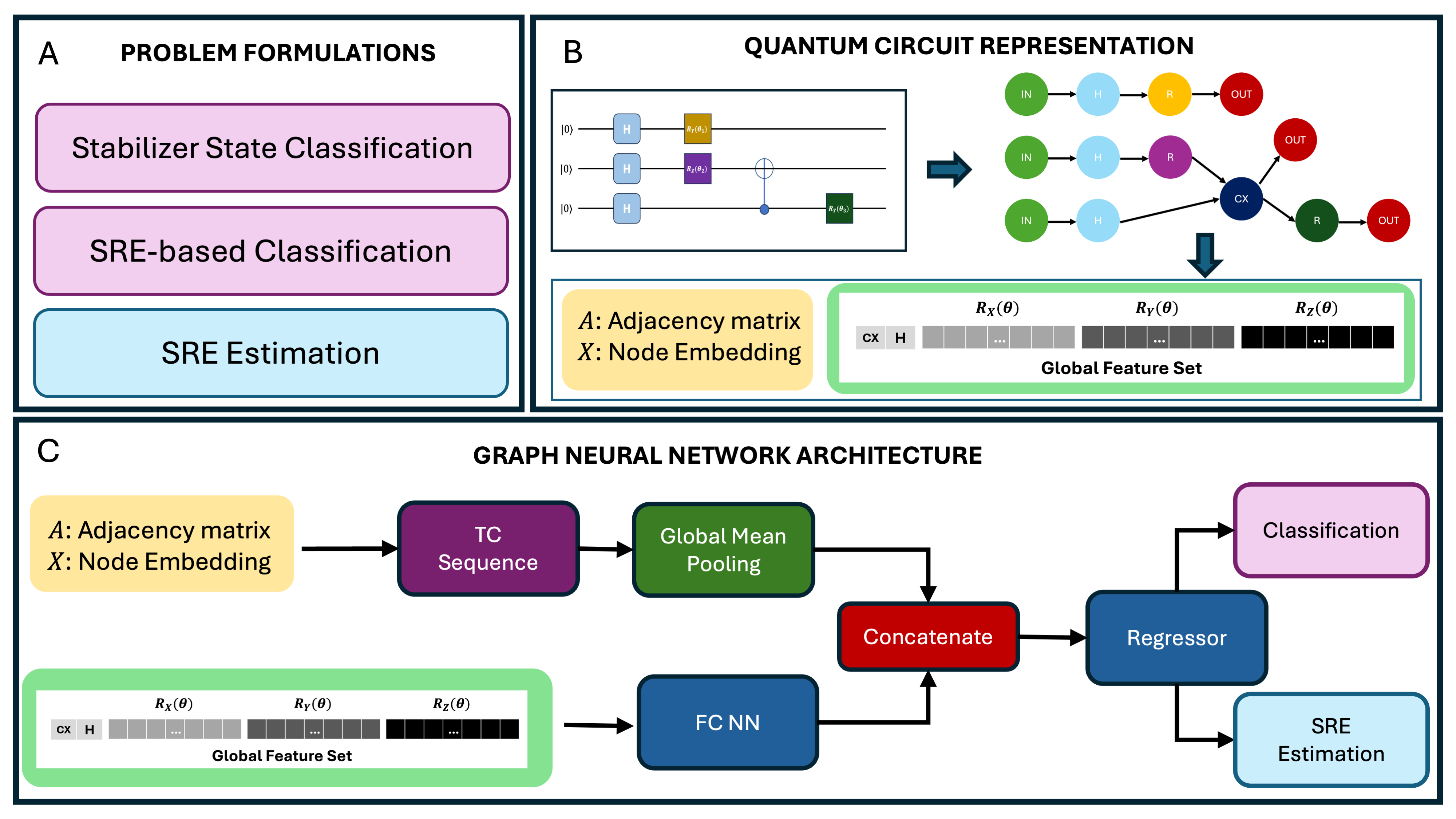}
    \caption{Overview of the GNN approach to nonstabilizerness estimation. Box A summarizes the three different problem formulations, box B illustrates the process to translate a quantum circuit in the graph-based representation, and box C illustrates the GNN architecture.}
    \label{fig:overview}
\end{figure}

This section describes the proposed Graph Neural Network (GNN) approach for nonstabilizerness estimation under both classification and regression problem formulations. All problem formulations share the same graph-based quantum circuit representation and GNN architecture. Figure~\ref{fig:overview} provides an overview of the whole pipeline. The proposed graph-based representation consists of two main components. The first, in yellow, is the directed acyclic graph representation of the circuit, which is encoded as an adjacency matrix together with node embeddings that represent quantum gates. The second component, in green, is the global feature set encoding the total gate count. These two components are processed independently, the first is passed through a sequence of Transformer convolutional layers, while the second through a fully connected neural network. The resulting latent representations are concatenated and passed through a regressor. Section~\ref{representation} and Section~\ref{gnn} describe the details of the encoding technique employed and the GNN architecture, respectively.
\subsection{Quantum Circuit Representation}\label{representation}

Quantum circuits can be naturally represented by directed acyclic graphs, where each node is a gate and each edge is a qubit acted upon by a gate~\cite{liao2024machine,aktar2024graph}. The graph passed in input to the GNN is encoded by two matrices, the adjacency matrix $A$ representing the graph structure, and the node feature matrix $X$, which contains the node embeddings. Since we deal with directed graphs, the adjacency matrix is not symmetric. Each node has an associated node embedding $x\in \mathbb{R}^{d}$. The size of the node embedding is $d=d_g+d_q+d_h$, as it stores the information regarding the gate type in $d_g$ components, the target qubit(s) upon which they are applied in $d_q$ components, and the specifics of the quantum backend in $d_h$ components,. The gate type and qubit indices are encoded using one-hot encoding, while the quantum backend vector is intrinsically real-valued. The one-hot encoding for the gate type requires $d_g=7$ bits, one for each gate type: input gate (qubit initialization), output gate (measurement), \texttt{CNOT}, \texttt{H}, \texttt{RX}, \texttt{RY}, and \texttt{RZ}. The one-hot encoding for the target qubit requires a number of bits equal to the number of qubits $d_q$ in the largest quantum circuit of the dataset. The hardware-dependent noise specifics of the gate applied require a real-valued vector of $d_h=7$ seven components, encoding the relaxation time $T^q_1$, the dephasing time $T^q_2$ and readout error $r_q$ on the qubit(s) $q$ on which the gate is applied, and the gate error $g_\epsilon$. Thus, in the stabilizer state classification and the SRE-based classification, where the datasets include quantum circuits with up to $25$ qubits the dimension of the node embedding is $d=7+25+0=32$. In the noiseless SRE estimation task, where circuit have up to $6$ qubits the embedding dimension is $d = 7 + 6 = 13$. In the noisy SRE estimation, we concatenate to the base node embedding of the noiseless case the $7$-dimensional vector encoding the specifics of the IBM Fake Oslo backend, resulting in $d = 13+7=20$. Figure~\ref{fig:overview} provides a comprehensive overview on the whole pipeline, including the architecture of the model.

\begin{figure}[!b]
    \centering
    \includegraphics[width=0.9\linewidth]{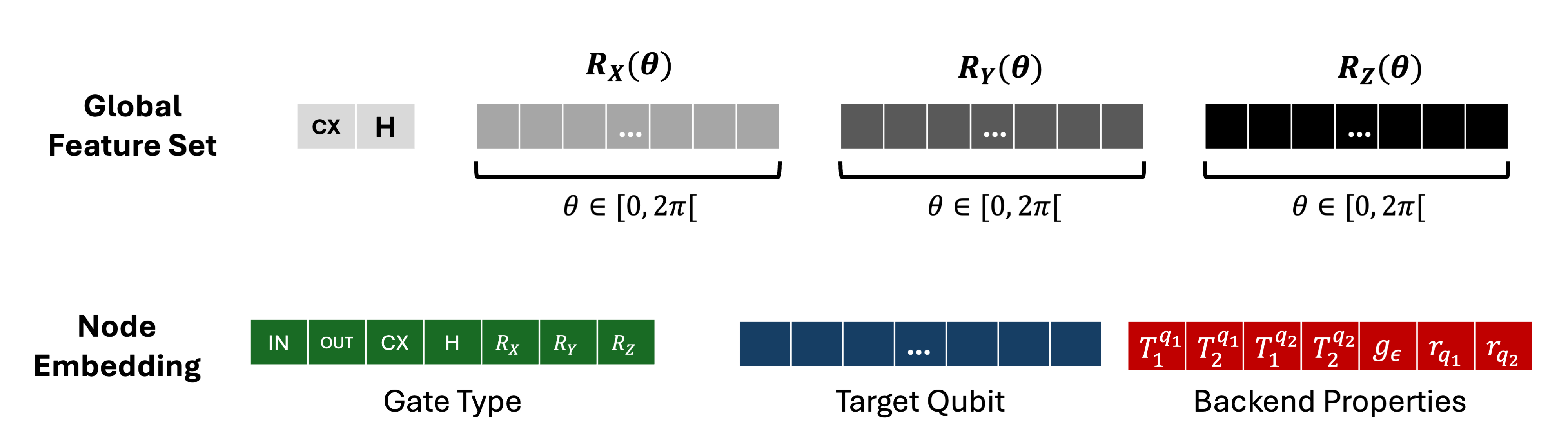}
    \caption{At the top, the global feature set, which is a vector in $\mathbb{R}^{d}$ encoding the gate count. The unparameterized gates are simply counted, while the parameterized gates are first divided into $50$ components, discretizing the angle parameter, and then counted. At the bottom, the (local) node embedding, which is a vector in $\mathbb{R}^{d}$ encoding gates, including their type, position, and specifics related to their implementation on hardware.  }
    \label{fig:features}
\end{figure}
The global feature set provided as input to the GNN model is a vector that encodes the gate count of quantum circuits. This representation has proven effective for supervised learning approaches to SRE estimation~\cite{lipardi2025study}. It consists of a vector of length $152$, where unparameterized gates are directly counted, whereas parameterized gates are discretized by dividing the interval $[0, 2\pi[$ of the angle parameter into fifty equally spaced bins. Figure~\ref{fig:features} illustrates both the global feature set and the node embeddings of the graph-based circuit representation.

The quantum circuit representation is exactly the same for all the three formulations of the nonstabilizerness estimation problem. However, in the noisy scenario, we had to take into account the universal gate set of the IBM Fake Oslo backend. Then, we adapted the encoding on the new gate set and trained the model on the transpiled quantum circuits.

\subsection{Graph Neural Network Architecture} \label{gnn}
The architecture of the Graph Neural Network employed in this work is inspired by previous works, where GNNs models are trained to predict the expressibility of quantum circuits~\cite{aktar2024graph} and to predict noise impact on circuit fidelity~\cite{wang2022torchquantum}. 

The model is divided into two main independent components.
The first component processes the graph representation, including the adjacency matrix $A$ and the node embeddings $X$, which are initialized as vectors encoding the gate type and target qubit(s), as described in the previous section.
These are processed by three Transformer Convolutional (TC) layers that aggregate local neighborhood information within the graph. A ReLU activation function is applied between each pair of consecutive TC layers to mitigate potential numerical issues arising from negative values~\cite{aktar2024graph}. Finally, the resulting feature vectors, obtained after applying the sequence of TC and ReLU layers, are aggregated using global mean pooling.
The second component is a fully connected artificial neural network that takes as input the global feature set, the vector encoding the gate counts of entire quantum circuits, as defined in the previous section.
Finally, the output vectors from the two components are concatenated and passed through a regressor consisting of three fully connected layers interleaved with two ReLU activation layers.

There are some differences between the GNN implementations for the classification and regression tasks. In the stabilizer state classification setting, we use the binary cross-entropy loss function, and the output of the GNN is passed through a sigmoid function with a threshold of $0.5$ to assign the class label of each quantum circuit. In contrast, for the SRE estimation problem, we employ the Huber loss function, and the final output of the GNN directly represents the predicted $M_2$ value of the input quantum circuit. Additionally, the circuit representations differ slightly across the various experimental configurations, as described in the previous section. 
Note that we employed a grid search to tune some hyperparameters of the GNN, such as the number of the TC layers, the hidden dimension of the fully connected neural networks, and the learning rate.

\section{Experimental Results}\label{results}
This section reports the experimental setup and results conducted to assess the proposed GNN approach on three different problem formulations of increasing difficulty.
Section~\ref{classification} focuses on the stabilizer state classification task, Section~\ref{exp_sre_classification} on the more challenging SRE-based classification task, and 
Section~\ref{exp_regression} on the SRE estimation task.
Each section outlines the training procedure and performance evaluation by assessing the generalization capability of the proposed GNN in predicting $M_2$ on quantum circuits with a higher number of qubits, a larger total gate count, and greater entanglement levels than those used for training.
Section~\ref{neural_networks} investigates the sources of improvement achieved by our GNN relative to prior SVR-based approaches. To this end, we evaluate the model’s performance following the ablation~\cite{cohen1988evaluation,newell1975tutorial} of the graph-based representation and the Transformer
layers that process it, resulting in a simplified fully connected neural network trained only on the global feature set.
Finally, Section~\ref{experiments_noise} extends the SRE estimation experiments to a noisy setting, considering transpiled quantum circuits labeled by simulating the noise of an IBM quantum backend.

\subsection{Stabilizer State Classification} \label{classification}
The experiments for training and testing our GNN model on the stabilizer state classification problem can be divided into two main parts. The first part focuses on quantum circuits with a fixed number of qubits, set to $18$, to allow a direct comparison with the previous approach~\cite{mello2025retrieving}. The second part extends the experiments to a broader dataset of quantum circuits with the number of qubits ranging from $2$ to $25$. 
\begin{figure}[!t]
    \centering
    \includegraphics[width=0.75\linewidth]{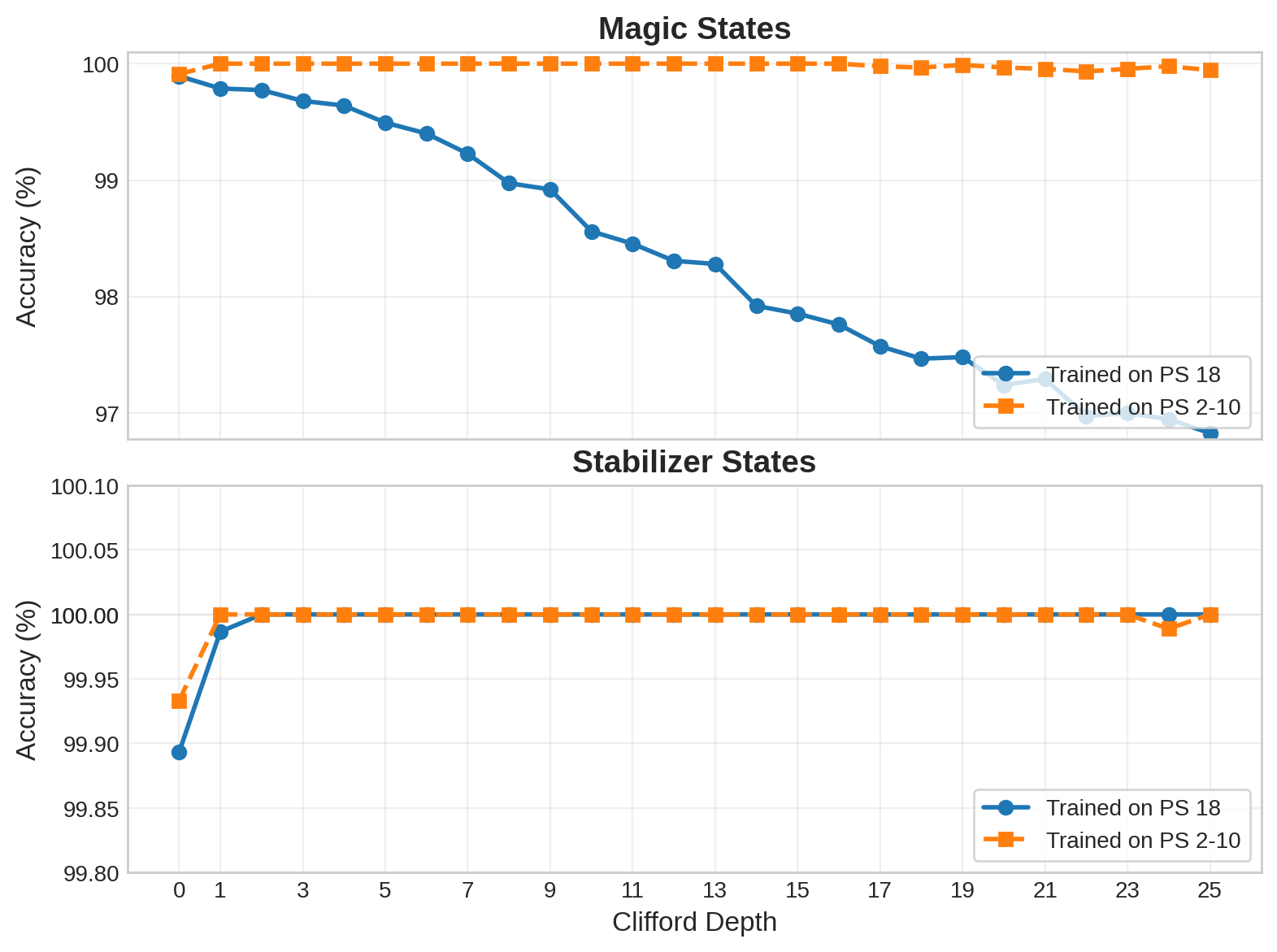}
    \caption{Stabilizer State Classification. Accuracy of the GNN model independently trained on two datasets of product states and evaluated on their corresponding Clifford-evolved datasets. The top plot shows the results for magic states, and the bottom plot shows those for stabilizer states. Results in blue correspond to the GNN trained on the $18$-qubit product state dataset (PS 18), while results in orange correspond to the GNN trained on product states with a number of qubits ranging from $2$ to $10$.}
    \label{fig:clifford_depth}
\end{figure}

In the first part of the experiments, GNN is trained on $10000$ out of the $15000$ circuits in PS 18, where both the training and test set are balanced between the two labels, stabilizer and magic states.
Our GNN achieves a classification accuracy of $99.98\%$ on the training set and $99.72\%$ on the test set. 
When evaluated on the Clifford-evolved circuits of the dataset CS 18, the GNN accuracy keeps stable for the classification of stabilizer states, while it decreases as a function of the Clifford depth for magic states, see the blue lines in Figure~\ref{fig:clifford_depth}. This phenomenon is also experienced by the CNN model. However, the classification accuracy obtained by the CNN decreases when evaluated on Clifford-evolved quantum states with Clifford depth ranging from $0$ to $8$. For magic states, it decreases from $100\%$ to approximately $90\%$, and for stabilizer states, from $100\%$ to $80\%$. In contrast, our GNN manages to maintain stable the classification accuracy for stabilizer states, and for magic states the accuracy only decreases from almost $100\%$ to $99\%$ at depth $8$. Furthermore, we extended the analysis up to a Clifford depth of $25$, where the accuracy obtained by the GNN on magic states decreases to $97\%$, while the accuracy on the stabilizer states is still stable to $100\%$. We refer the reader to Figure~\ref{fig:clifford_depth}.

\begin{figure}[!hb]
    \centering
    \begin{subfigure}[t]{0.49\textwidth}
        \centering
        \includegraphics[width=\linewidth]{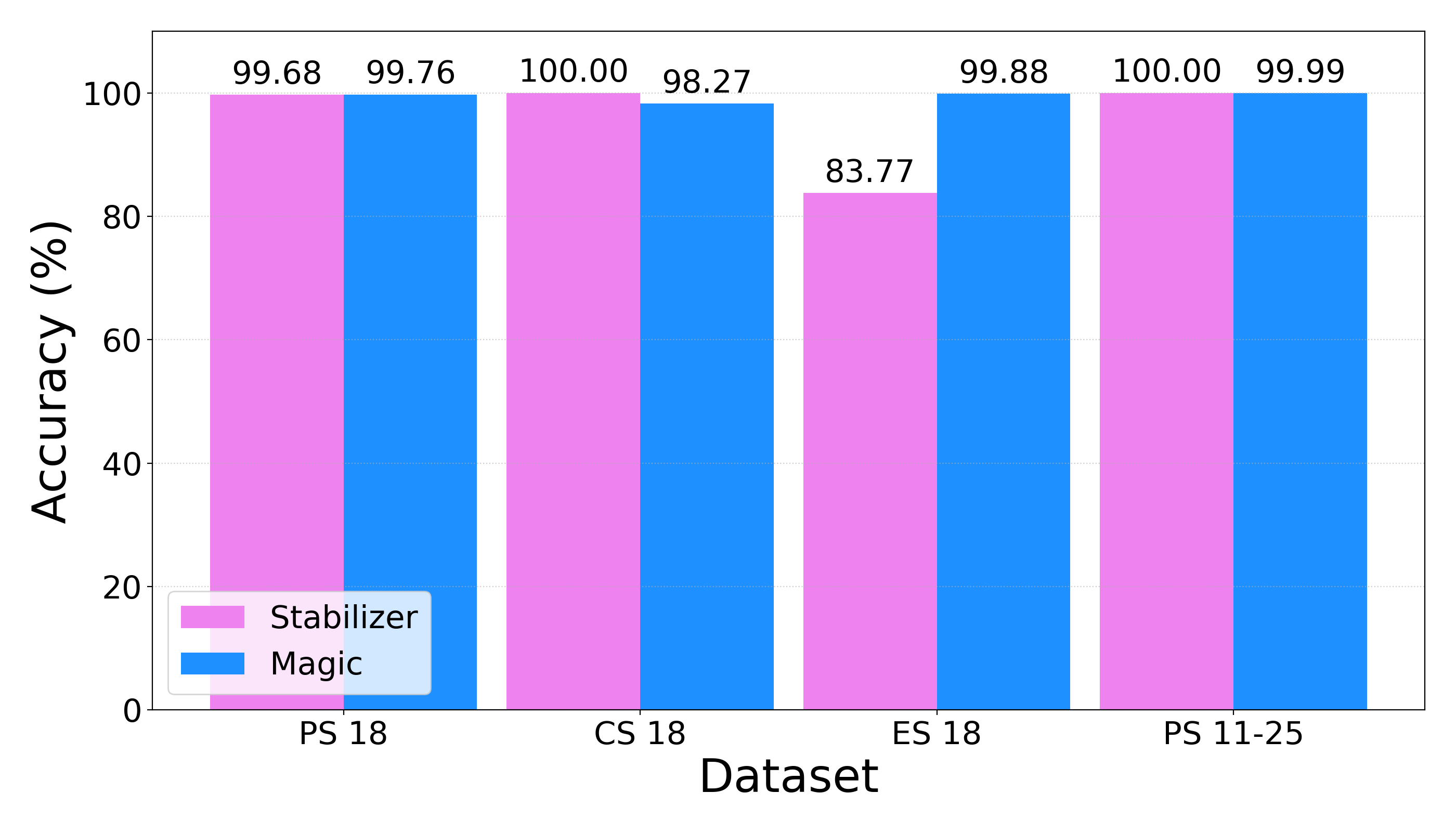}
        \caption{Trained on PS 18}
        \label{fig:product_18}
    \end{subfigure}
    \begin{subfigure}[t]{0.49\textwidth}
        \centering
        \includegraphics[width=\linewidth]{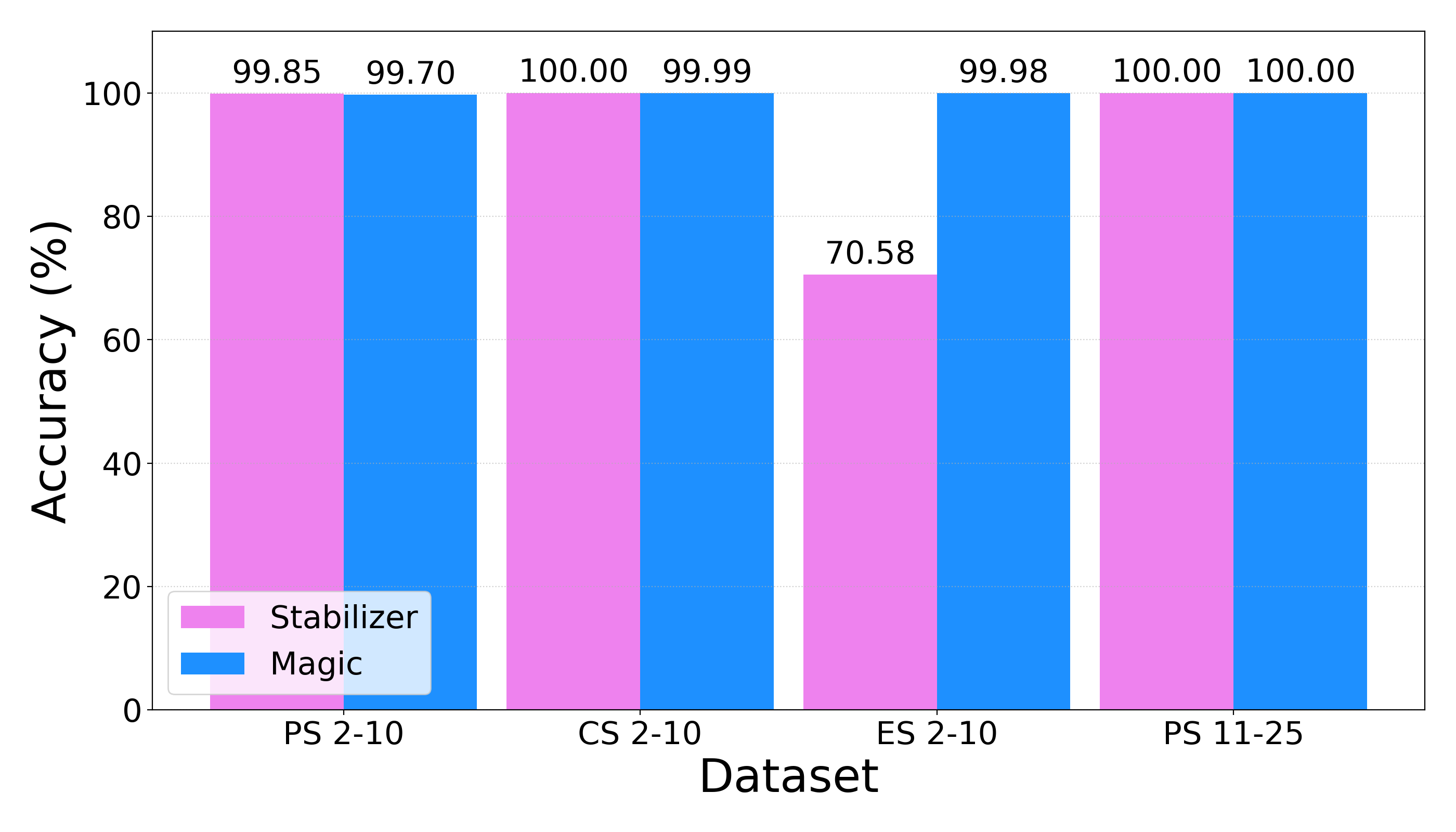}
        \caption{Trained on PS 2-10}
        \label{fig:product_2_10}
    \end{subfigure}
    \caption{Evaluation of the generalization capabilities of the GNN model, in terms of classification accuracy, across different datasets for both stabilizer states (purple) and magic states (blue).}
    \label{fig:accuracy_classification}
\end{figure}

Additionally, we extended the evaluation performance of the GNN models to other types of out-of-distribution instances. These include the circuits of the dataset ES 18, which keep the same qubit number but with a substantially different structure, due to a high number of CNOT gates injected randomly throughout the circuits, and the circuits in the dataset PS 11-25  which are also product states as those used during training but with different number of qubits. Figure~\ref{fig:product_18} illustrates the accuracy obtained by the model across the different datasets, each corresponding to testing GNN on a distinct generalization task. Specifically, PS 18 tests generalization to unseen circuits within the same dataset used for training, CS 18 evaluates generalization to circuits evolved under random Clifford gates, and ES 18 assesses generalization to circuits with higher entanglement and different structural properties.

In the second part, the GNN model is trained on circuits with low qubit numbers, specifically from $2$ to $10$. The training and test sets are split in a $70$ to $30$ ratio and kept balanced between the two classes. In this case, our GNN achieves a classification accuracy of $99.98\%$ on the training set and $99.78\%$ on the test set. Despite the previous case of $18$-qubit circuits, the GNN manages to maintain approximately a perfect classification accuracy on both stabilizer and magic states on the subset of circuits with number of qubits between $2$ and $10$ of the dataset CS 2-25, as indicated by the orange line in Figure~\ref{fig:clifford_depth}. The same GNN model is subsequently evaluated on different generalization tasks, following the same methodology as in the previous case. Figure~\ref{fig:product_2_10} illustrates the classification accuracy for both stabilizer and magic states across various datasets. Remarkably, the model successfully classifies with high accuracy the product states with higher numbers of qubits in the PS 11–25 dataset, their Clifford-evolved counterparts in the CS 2–10 dataset, and the circuit with different structure and higher level of entanglement in ES 2-10. These experimental results disclose the strong generalization performances achieved by the GNN model in the stabilizer state classification task.

Furthermore, we report in Section~\ref{ssc_complementary} an analysis on the distribution of the misclassified instances in terms of their $M_2$ density.

\subsection{SRE-based Classification} \label{exp_sre_classification}
The experiments for the SRE-based classification task follow the same procedure described for the stabilizer state classification. In this experimental setup, the GNN model is trained on product-state circuits with qubit numbers up to $10$ (dataset PS 2–10), using a train–test split ratio of $70$ to $30$. The generalization performance is then evaluated on three different scenarios: unseen circuits from the test set, product states with a higher number of qubits (dataset PS 11–25), and Clifford-evolved circuits with depths up to $25$ (dataset CS 2–10).
\begin{figure}[!hb]
    \centering
    \begin{subfigure}[t]{0.49\textwidth}
        \centering
        \includegraphics[width=\linewidth]{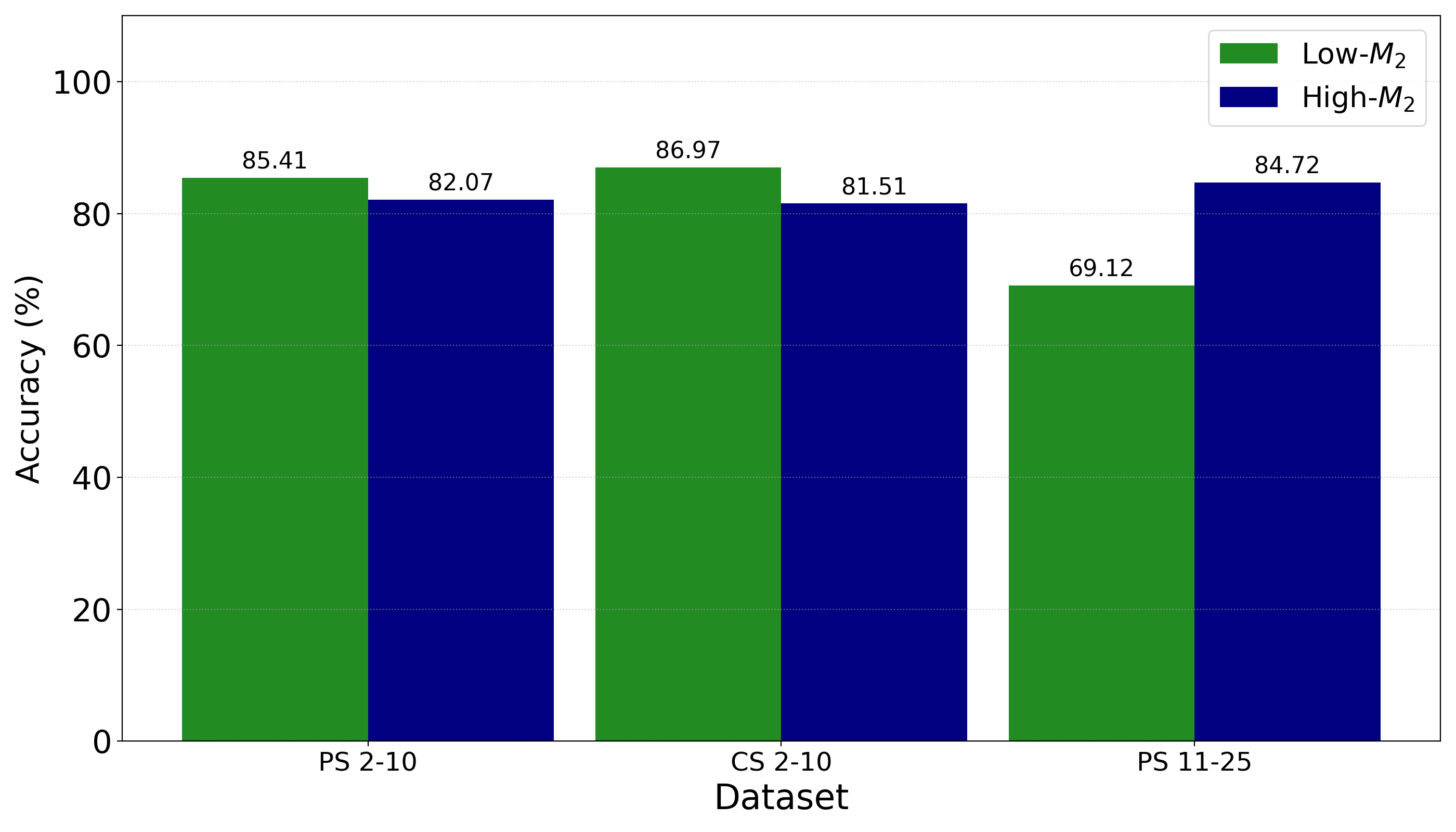}
        \caption{Generalization across different scenarios}
        \label{fig:sre-based_accuracy}
    \end{subfigure}
    \begin{subfigure}[t]{0.49\textwidth}
        \centering
        \includegraphics[width=\linewidth]{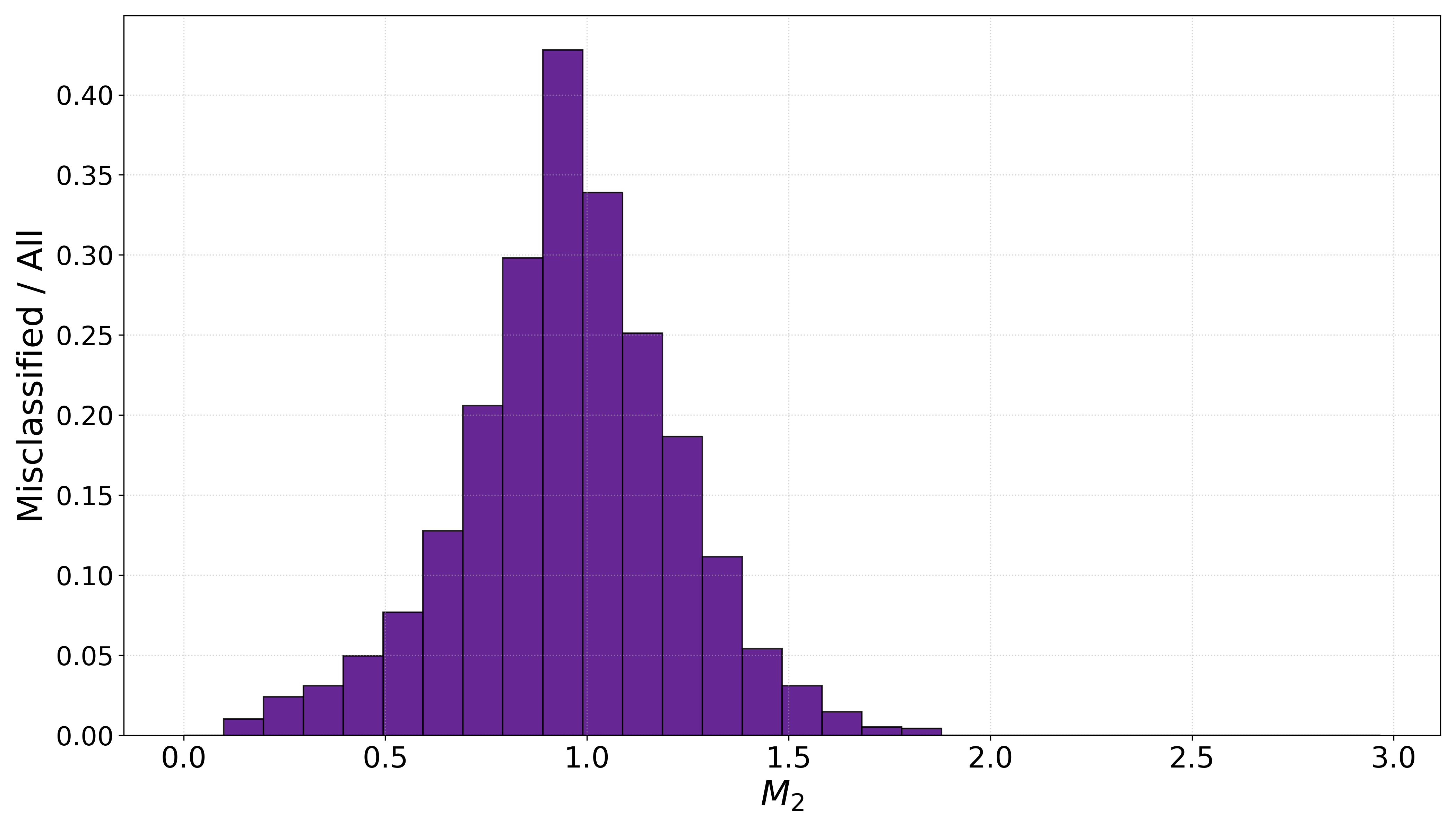}
        \caption{Performance dependence on $M_2$}
        \label{fig:results_sre_dependence}
    \end{subfigure}
    \caption{Evaluation of the GNN model for the SRE-based classification. On the left, the classification accuracy to evaluate the generalization performances across different scenarios. On the right, a study on the performance of the GNN model depending on $M_2$.}
    \label{fig:result_sre-based}
\end{figure}

Figure~\ref{fig:result_sre-based} presents the experimental results of the GNN model trained for the SRE-based classification task on the PS 2–10 dataset. Figure~\ref{fig:sre-based_accuracy} reports the classification accuracies across the three evaluation scenarios, for both low-$M_2$ circuits (in green) and high-$M_2$ circuits (in blue). Even for this more challenging classification task, the GNN achieves consistently high accuracy. Across all three scenarios, the accuracies remain above $80\%$, except for the case of low-$M_2$ circuits in the PS 11–25 dataset. In this latter dataset, the $M_2$ values differ substantially from those observed during training, with the maximum increasing from $3$ to $5$ and the median shifting from $0.90$ to $1.61$, as shown in the Figure~\ref{fig:sre_based_data_distribution}.
Figure~\ref{fig:results_sre_dependence} shows the ratio of circuits misclassified by the GNN to the total number of circuits in the dataset, plotted as a function of their $M_2$ values. The plot exhibits a Gaussian-like distribution centered around the $M_2$ threshold ($0.90$). This behavior is expected, as the classification task becomes increasingly challenging for circuits with $M_2$ values close to the threshold, where distinguishing between low- and high-$M_2$ circuits is inherently more difficult. We note that in the previous section we provide a study on the performance depending on $m_2$, as it was a specific case of fixed qubit number addressed to compare with previous work. In this more general case, is the $M_2$ value of the whole circuit that matter.

\begin{figure}[!b]
    \centering
    \includegraphics[width=0.75\linewidth]{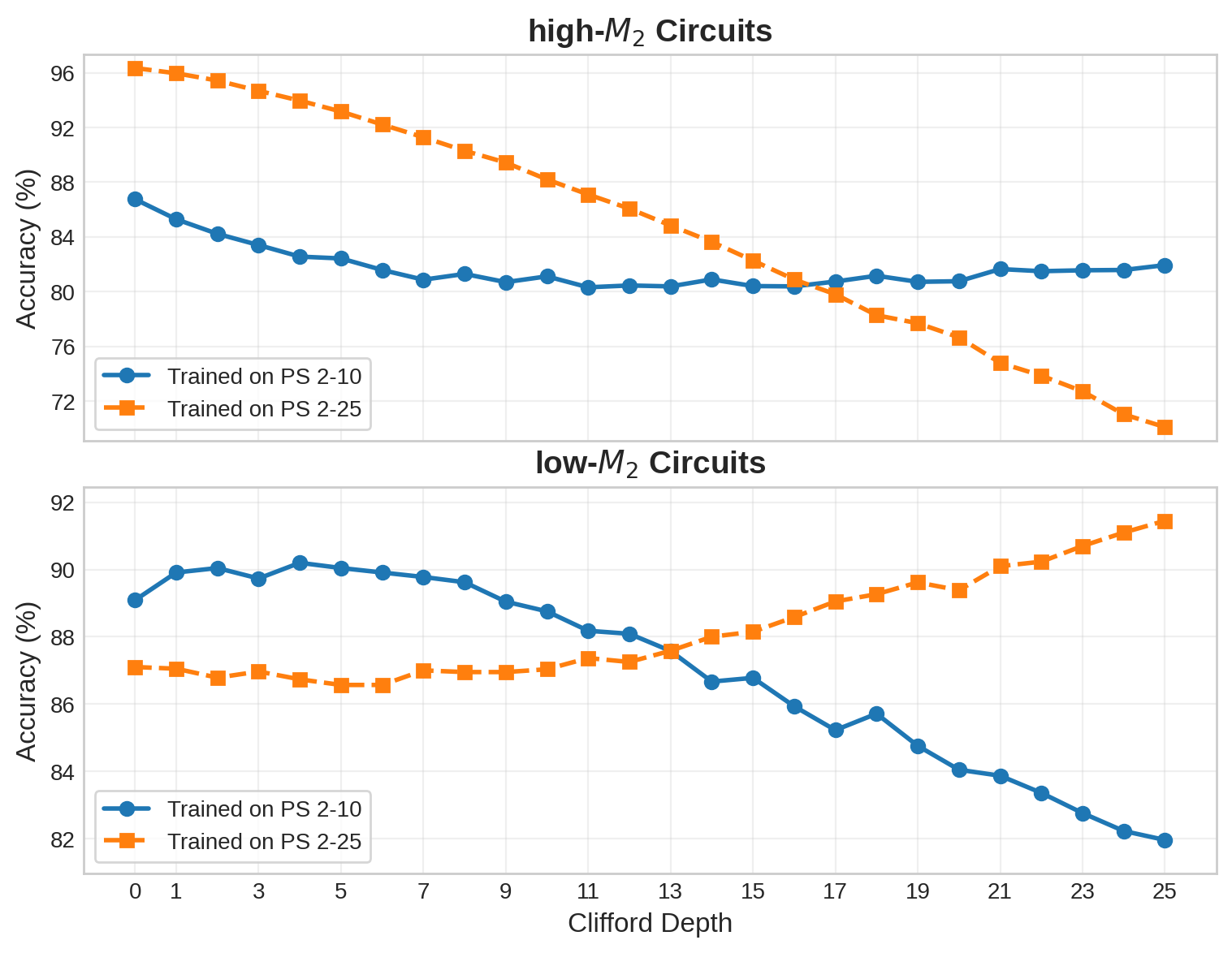}
    \caption{Accuracy of the GNN model for the SRE-based classification task. The GNN is independently trained on two datasets of product states and evaluated on their corresponding Clifford-evolved datasets.}
    \label{fig:clifford_depth_sre}
\end{figure}

Figure~\ref{fig:clifford_depth_sre} provides a detailed analysis of the classification accuracy achieved by the GNN on the Clifford-evolved datasets as a function of Clifford depth. For this analysis, we additionally trained the GNN on the PS 2–25 dataset to study how the model performance scales with the problem size. The blue curves correspond to the accuracies obtained by the GNN trained on PS 2–10 and evaluated on CS 2–10, while the orange curves represent the results of the GNN trained on the larger PS 2–25 dataset and evaluated on CS 2–25. The top plot reports the accuracies for high-$M_2$ quantum circuits, whereas the bottom plot shows the accuracies for low-$M_2$ circuits.
Interestingly, scaling up the problem reveals a distinct behavior. For high-$M_2$ circuits, the accuracy achieved on smaller systems (2–10 qubits) remains more robust under Clifford evolution compared to the more challenging case with 11–25 qubits. Conversely, for low-$M_2$ circuits, while the accuracy decreases with Clifford depth in the smaller-qubit case, it improves for the larger one.
It is important to note that all circuits in the datasets contain both Clifford and non-Clifford gates. As the Clifford depth increases, not only does the circuit structure change, but also the ratio of Clifford to non-Clifford gates grows. Furthermore, the larger PS 2–25 dataset offers greater variance and a larger sample size, which contribute to the model’s improved generalization behavior. The improvement in the accuracy might be the sign that our GNN model managed to capture this structural feature.

\subsection{Stabilizer R\'enyi Entropy Estimation} \label{exp_regression}
The experiments for training and testing our GNN model for SRE estimation are aligned with the setting proposed in~\cite{lipardi2025study}. To assess the generalization capabilities of our GNN approach, we test its performance on out-of-distribution instances. The experiments can be divided into two main parts. The first part is designed to evaluate the generalization capabilities of GNN in terms of number of qubits, while the second part in terms of gate counts.
We refer to these experiments as \textit{extrapolation} in qubit number and in gate count, respectively. \rev{To further assess the performance of the proposed GNN, we include additional analysis in the appendices~\ref{generalization_to_structured_circuits} and ~\ref{appendix_analytic}.
Specifically,} Appendix~\ref{generalization_to_structured_circuits} describes additional experiments where the GNN trained on unstructured random quantum circuits is evaluated on well-known structured circuits, based on the TIM and Heisenberg model. \rev{Appendix~\ref{appendix_analytic} presents a benchmark comparison between the GNN predictions and estimates derived from the analytic expression introduced in Equation (13) of Ref.~\cite{leone2022stabilizer}.}
\begin{figure}[!ht]
    \centering
    \includegraphics[width=0.95\linewidth]{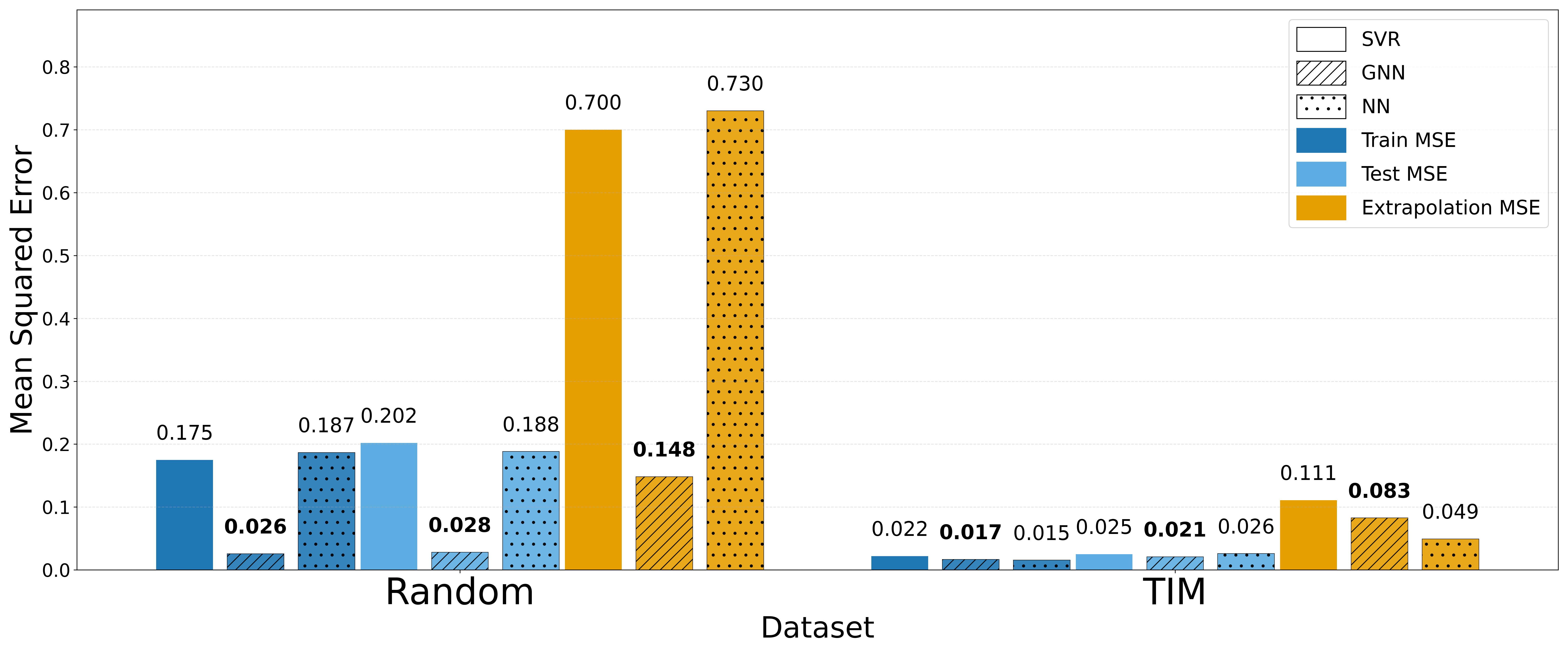}
    \caption{Performance comparison between GNN, SVR (hatched), and NNs (dotted) in the setup of extrapolation per qubit number.}
    \label{gnn_extrapolation_qubits}
\end{figure}

We train our GNN independently on the two quantum circuits datasets, RQC and TIM. In all experiments we split training and test set in $80$-$20$. Note that, to ensure statistical robustness, each experiment is repeated 10 times with different random seeds, and the reported results correspond to the average results over these ten independent runs. 
Figure~\ref{gnn_extrapolation_qubits} summarizes the experiments designed to test the extrapolation in the qubit number. Here, the models are trained on the subset of quantum circuits with qubit numbers from $2$ to $5$, and then evaluated on $6$-qubit circuits. The difference between the MSE values obtained by the SVR on the training and test set suggests the presence of overfitting. This effect is observed in both datasets. In the RQC dataset, the MSE increases from $0.175$ on the training set to $0.202$ on the test set, while in the TIM dataset, the MSE increases from $0.22$ to $0.25$. As a result, the generalization performances are limited as the MSE increases of approximately of a factor $3.5$ in the RQC dataset and $4.5$ in the TIM dataset. In contrast, the GNN model significantly improves upon the results obtained with SVR, particularly in terms of Mean Squared Error (MSE) on the extrapolation set. Specifically, GNN reduces the extrapolation MSE upon SVR by the $79\%$ on RQC, and by $26\%$ on TIM. The highest improvement is achieved on the RQC dataset, where the GNN also reduces the gap between the MSE values obtained on the training and test set. We refer the reader to Appendix~\ref{sre_complementary}, where we investigate the motivation of the improved MSE obtained by NN compared to GNN on the TIM dataset (Figure~\ref{gnn_extrapolation_qubits}). Appendix~\ref{qubit_extrapolation_appendix} presents additional experiments evaluating the extrapolation capability of the GNN to circuits with up to $8$ qubits.
\begin{figure}[!b]
    \centering
    \includegraphics[width=0.95\linewidth]{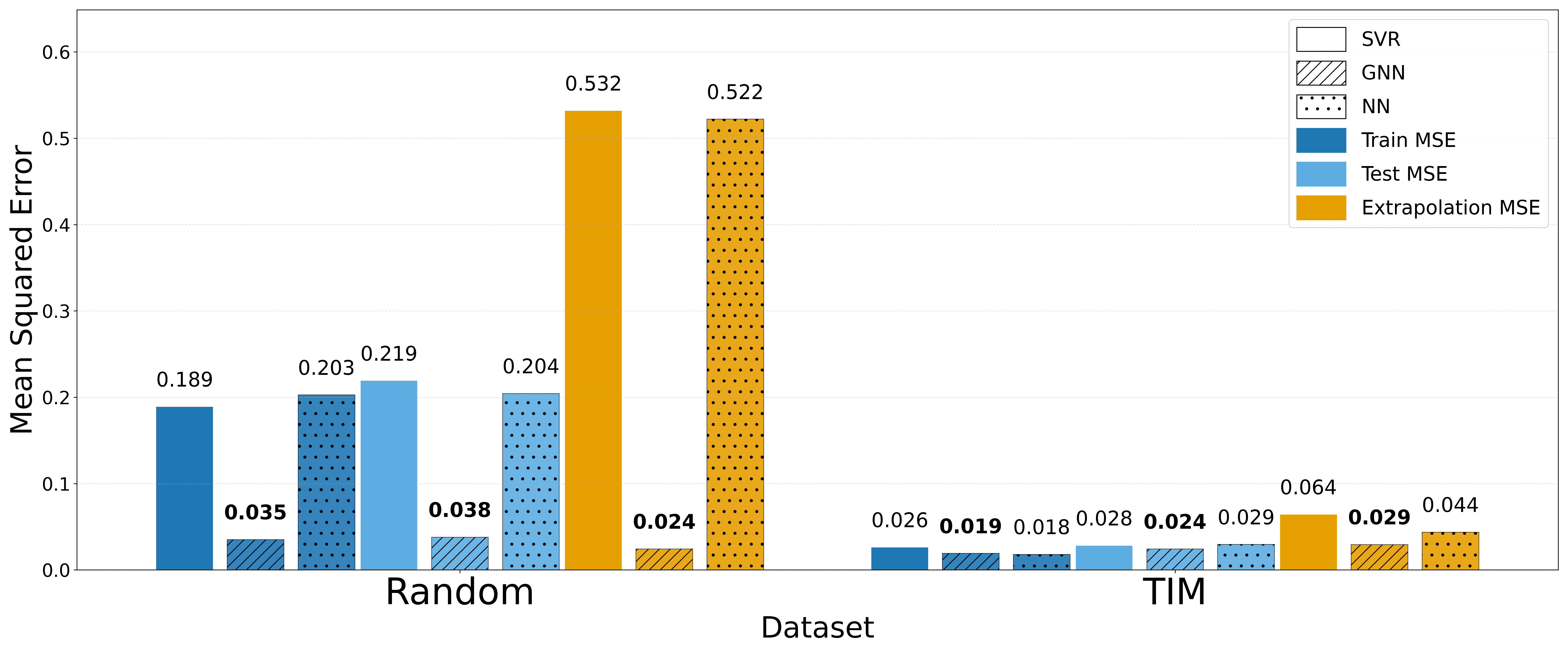}
    \caption{Performance comparison between GNN, SVR (hatched), and NNs (dotted) in the setup of extrapolation per gate count.}
    \label{gnn_extrapolation_depth}
\end{figure}

Figure~\ref{gnn_extrapolation_depth} summarizes the experiments designed to test the extrapolation in gate counts. Here, the models are trained on the subset of circuits with gate counts restricted to the range $[0,79]$ for the RQC dataset and evaluated on circuits with gate counts ranging in $[80,99]$. This approach is similarly applied to the TIM dataset, where the model is trained on circuits with trotter steps from $1$ to $4$ and tested on trotter step $5$. The difference between the MSE values obtained by the SVR on the training and test sets indicates a slight overfitting effect. This behavior is more evident on the RQC dataset, where the MSE increases from $0.189$ in the training set to $0.219$ in the test set. In contrast, for the TIM dataset, the MSE increases only from $0.026$ to $0.028$, resulting in a slightly better generalization performance compared to the RQC dataset. In this setting, our GNN demonstrates strong generalization performance. The gap between training and test MSE observed in the SVR model for the RQC dataset is reduced, and concurrently the extrapolation MSE improves by $95\%$, from $0.532$ to $0.025$. Moreover, the GNN achieves an extrapolation MSE of $0.028$, which improves the SVR model by $56\%$.


\subsection{The Importance of the Graph Representation}\label{neural_networks}
Given the improved performance of the GNN compared to the SVR model, it is important to investigate the source of this advantage. We identify two main differences between the GNN and SVR approaches. First, the GNN employs a graph-based representation combined with Transformer Convolutional layers, which aggregate local information from neighboring nodes. Second, the GNN leverages a deep learning architecture to process the global feature set, which is the same set used in the SVR model~\cite{lipardi2025study}.

To examine the importance of the graph representation, we perform experiments by isolating the fully-connected neural network from the upper part of the model shown in Figure~\ref{fig:overview}. This is a common process in artificial intelligence, known as ablation, where an individual component is obscured in order to evaluate its contributions to the whole complex models \cite{cohen1988evaluation,newell1975tutorial}. In this setup, we disabled the Transformers Convolutional responsible for processing the graph-structured data. Consequently, we train only the fully connected neural network (NNs) on the global feature set. The dotted bars in Figure~\ref{gnn_extrapolation_qubits} and Figure~\ref{gnn_extrapolation_depth} show the results obtained in this experimental setup for the extrapolation on qubit number and gate count, respectively. Table~\ref{tab:extrapolation-results} compares directly the extrapolation performances of GNN an NN with respect to the SVR model~\cite{lipardi2025study}.
\begin{table}[!ht]
\centering
\caption{Relative performance of the GNN and NN models compared to the SVR baseline~\cite{lipardi2025study}, in terms of the extrapolation MSE.}
\label{tab:extrapolation-results}
\begin{minipage}{0.45\textwidth}
\centering
\caption*{Extrapolation on Qubit Number}
\begin{tabular}{@{}lcc@{}}
\toprule
Dataset & GNN & NN \\
\midrule
RQC & -78.8\% & +4.3\% \\
TIM & -25.6\% & -55.6\% \\
\bottomrule
\end{tabular}
\end{minipage}
\hspace{0.025\textwidth} 
\begin{minipage}{0.45\textwidth}
\centering
\caption*{Extrapolation on Gate Counts}
\begin{tabular}{@{}lcc@{}}
\toprule
Dataset & GNN & NN \\
\midrule
RQC & -95.3\% & -1.9\% \\
TIM & -55.6\% & -31.6\% \\
\bottomrule
\end{tabular}
\end{minipage}
\end{table}

The experiments show that the graph representation has a significant impact on the SRE estimation problem in the RQC dataset. In the case of extrapolation on the qubit number, the $79\%$ improvement in MSE over SVR completely vanishes when training only a neural network on the global feature set, and performance even degrades, becoming $4\%$ worse than SVR. Similarly, for extrapolation on gate counts, the $95\%$ improvement in GNN vanishes to approximately $2\%$. However, on the TIM dataset, the impact of the graph representation is different. While NN loses nearly half of the advantage gained by the GNN (achieving a $33\%$ improvement compared to the $56\%$ for the GNN) in extrapolating on gate counts, the NN achieves a significantly better result than the GNN when extrapolating on the qubit number ($56\%$ versus $26\%$ for the GNN). These experimental results suggest that the circuit structure has the highest impact on the datasets where there is more diversity. In this regards, the RQC dataset is composed of random quantum circuits with different structure, while the TIM dataset has a structure of a fixed sequence of gates that repeats on different combination of qubits and with the diversity hidden in the angle parameters of the rotation gates.

\subsection{Quantum Noise Effect on Stabilizer R\'enyi Entropy Estimation} \label{experiments_noise}
One of the main motivations for adopting the Stabilizer R\'enyi Entropy (SRE) as a measure of nonstabilizerness in this work is its suitability for experiments on real quantum devices~\cite{oliviero2022measuring,niroula2024phase}. For this reason, in this section we assess the impact of quantum noise on the predictions of our GNN models \rev{for both the generalized SRE for mixed states $\tilde{M_2}$ and the witness of magic $\mathcal{W}_2$}. The RQC Noisy and TIM Noisy datasets have been generated specifically for this task. In these datasets, all quantum circuits are labeled with their $\tilde{M_2}$ \rev{and $\mathcal{W}_2$} values, according to Equations~\ref{sre_mixedstates} \rev{and~\ref{witness}}, obtained by simulating quantum noise using the IBM Fake Oslo backend.
\begin{figure}[!t]
    \centering
    \begin{subfigure}[t]{0.49\textwidth}
        \centering
        \includegraphics[width=\linewidth]{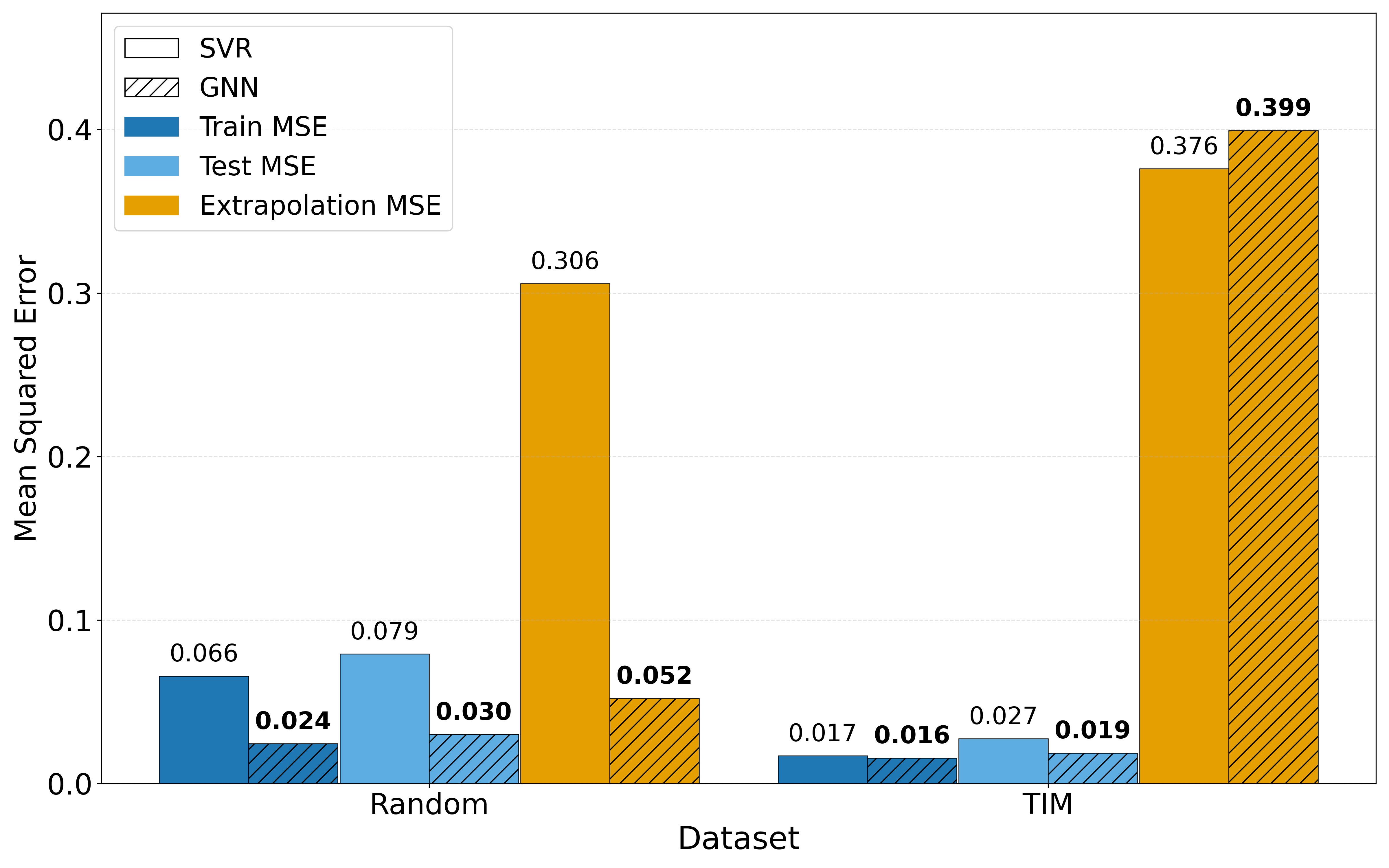}
        \caption{$\tilde{M}_2$: Extrapolation on qubit number}
        \label{fig:noise_extrapolation_qubit}
    \end{subfigure}
    \begin{subfigure}[t]{0.49\textwidth}
        \centering
        \includegraphics[width=\linewidth]{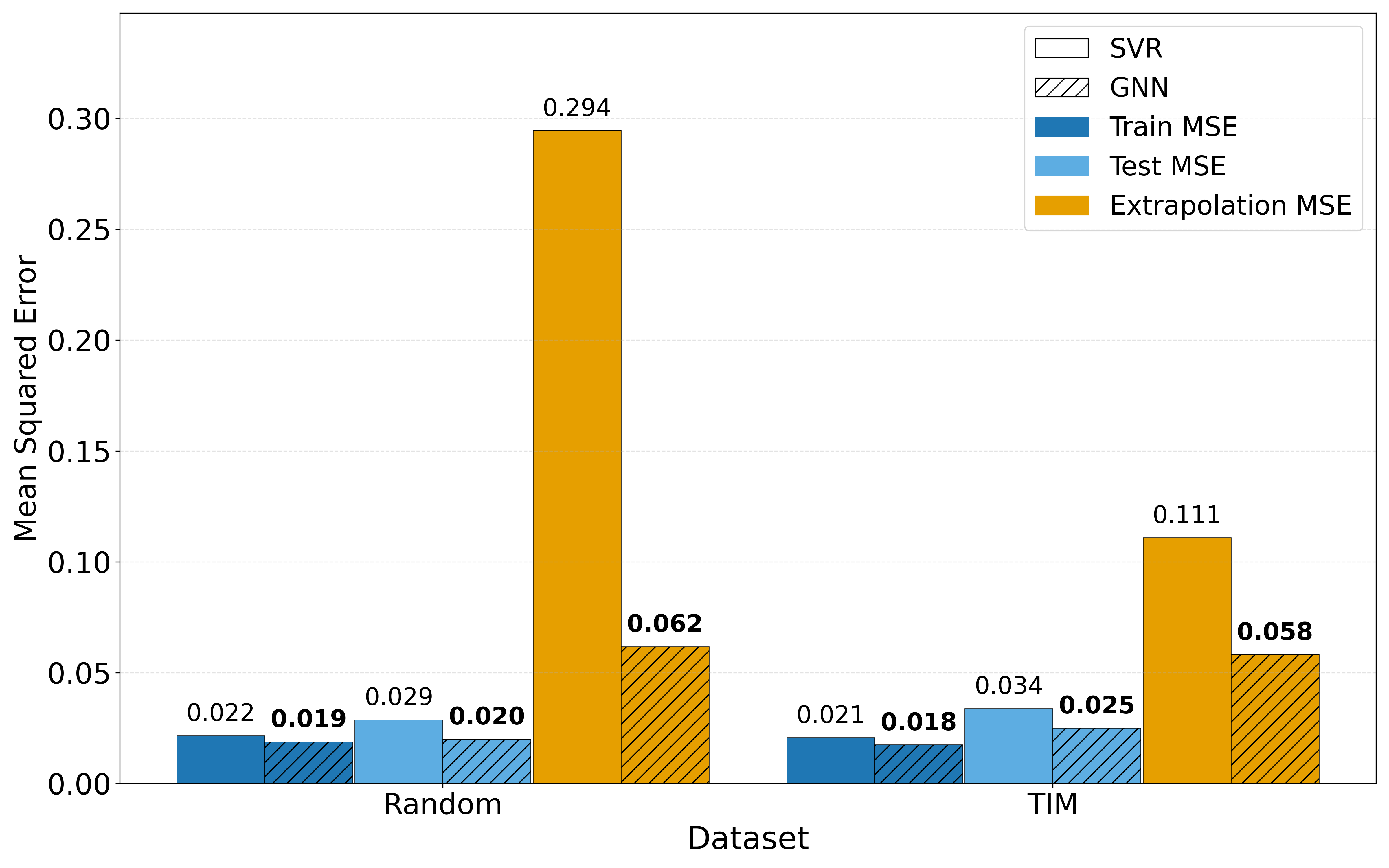}
        \caption{$\tilde{M}_2$: Extrapolation on circuit depth}
        \label{fig:noise_extrapolation_depth}
    \end{subfigure}
    \par\medskip
    \begin{subfigure}[t]{0.49\textwidth}
        \centering
        \includegraphics[width=\linewidth]{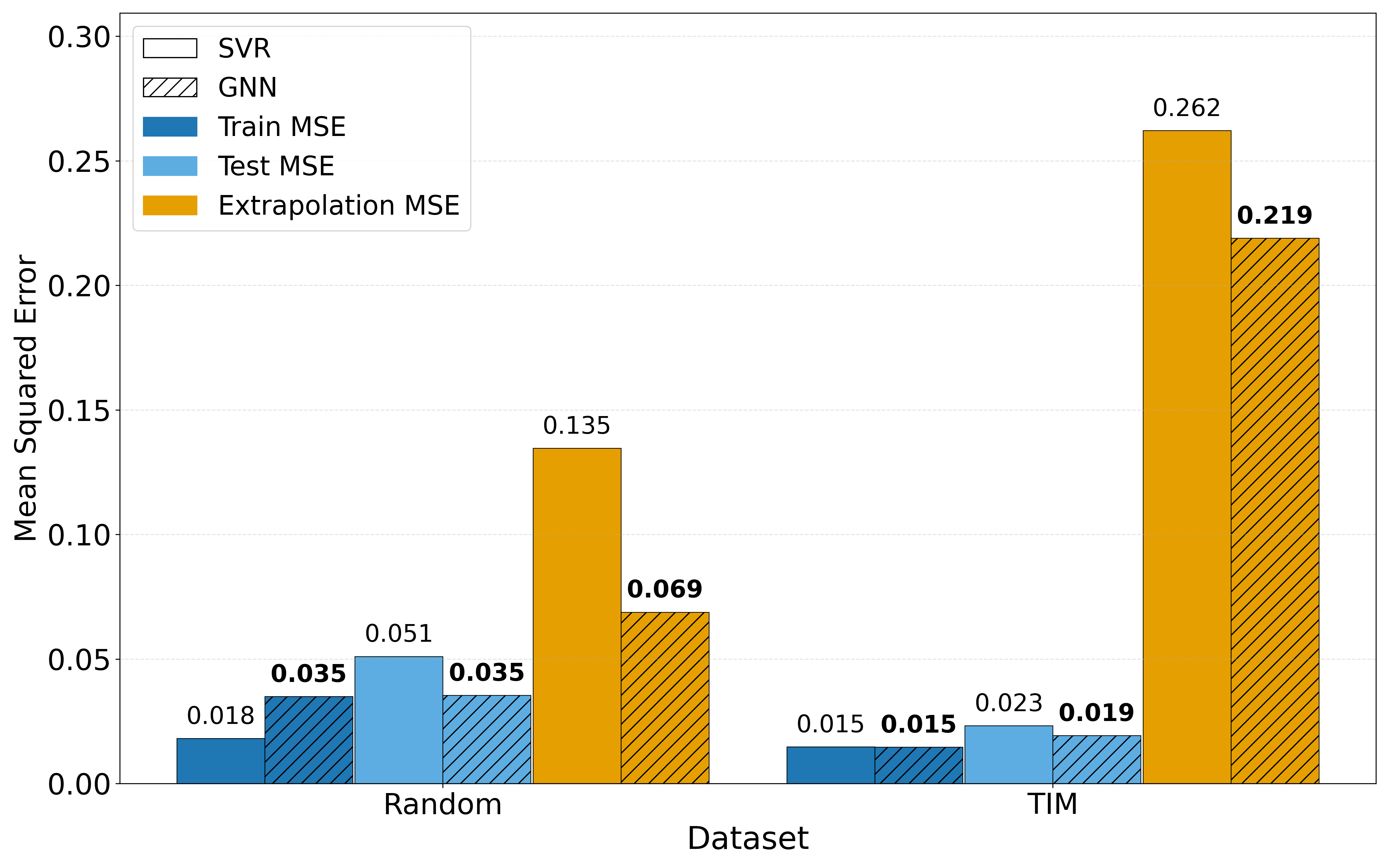}
        \caption{$\mathcal{W}_2$: Extrapolation on qubit number}
        \label{fig:noise_extrapolation_c}
    \end{subfigure}
    \hfill
    \begin{subfigure}[t]{0.49\textwidth}
        \centering
        \includegraphics[width=\linewidth]{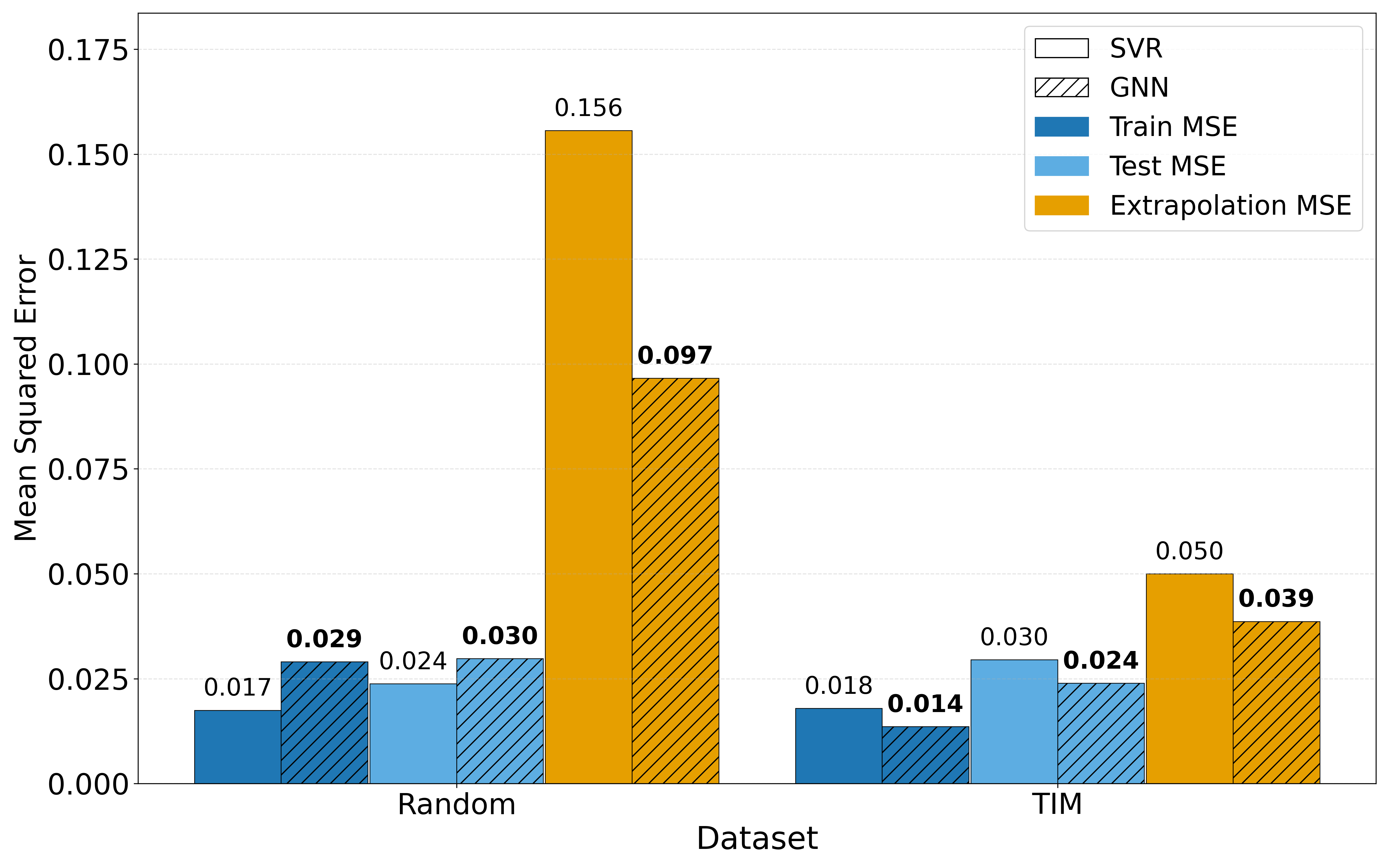}
        \caption{$\mathcal{W}_2$: Extrapolation on circuit depth}
        \label{fig:noise_extrapolation_d}
    \end{subfigure}
    \caption{Comparison between GNN and SVR (hatched) in terms of MSE on the RQC Noisy and the TIM Noisy datasets. }
    \label{fig:experiment_noise}
\end{figure}

The experiments in this setup follow exactly the same procedure as those described in Section~\ref{exp_regression}. 
Figure~\ref{fig:experiment_noise} reports the mean squared error (MSE) values for the extrapolation experiments with respect to the qubit number (Figures~\ref{fig:noise_extrapolation_qubit} \rev{and~\ref{fig:noise_extrapolation_c}}) and the gate count (Figures~\ref{fig:noise_extrapolation_depth} \rev{and~\ref{fig:noise_extrapolation_d}}). The SVR model was trained following the procedure described in~\cite{lipardi2025study}, including a grid search over the hyperparameters.

\rev{The GNN model improves the SVR predictions also in the noisy setup for both $\tilde{M_2}$ and $\mathcal{W}_2$, due to the local embedding of the circuit representation, which incorporates backend properties (see Figure~\ref{fig:features}). In line with the results obtained on the noiseless scenario in Section~\ref{exp_regression}, the GNN provides better improvements on the more general random quantum circuit dataset. These experiments further highlight the broad applicability of the proposed GNN framework for predicting different measures of nonstabilizerness, when suitably trained for the target measure.}

We note that the noisy RQC and TIM datasets are smaller in size compared to their respective noiseless counterparts, given the high computational time required for the simulation. For this reason, a comparison of the results between the model predictions obtained in the noisy and noiseless scenarios has to be interpreted with caution, as part of the performance gap may be attributed to the reduced training set size rather than the effect of noise itself.

\section{Conclusions}\label{conclusions}
This article proposed a Graph Neural Network (GNN) approach for addressing the nonstabilizerness estimation problem in the supervised learning setting through three formulations of increasing difficulty. 
The first is the stabilizer state classification, which is defined independently of any specific nonstabilizerness measure. This formulation serves primarily as a benchmark to establish a consistent data generation procedure and to perform an initial assessment of the GNN model on an easier problem. The second is the SRE-based classification defined on the stabilizer R\'enyi entropy (SRE). This formulation represents a more challenging and physically motivated task, as it directly connects to identifying stabilizer states based on their SRE measured on a real quantum device. Finally, the SRE estimation, is the more general regression task, which aims to predict the SRE of quantum circuits as a quantitative measure of nonstabilizerness.

In the first two classification tasks, the GNN model is trained on product-state quantum circuits. Compared to previous CNN-based approaches~\cite{mello2025retrieving}, the proposed GNN model achieves more robust generalization performance on circuits evolved under Clifford operations. Since the \texttt{CNOT} is a Clifford gate, Clifford-evolved circuits allow to test the generalization capabilities to entangled states. Our proposed data generation procedure includes not only stabilizer and highly magic states but also quantum states with low SRE values. This diversity is a fundamental feature of the dataset, as it prevents the problem from being defined on states that are easily separable. Additionally, we extended the analysis to circuits with different structure and high number of CNOT gates, as well as on circuits with higher number of qubits, where the GNN also exhibits robust performances.
For the regression task, our GNN model provides more accurate SRE predictions than the Support Vector Regressor~\cite{lipardi2025study}, especially in terms of generalization to larger, previously unseen quantum circuits. By ablation of the graph representation and the GNN components that process the graph data, we show that the graph representation improves the model predictions. The impact of the graph representation is significantly relevant on random quantum circuit where circuit structures are quite different.
Moreover, the experimental results show that the proposed GNN is a promising approach to predict SRE values measured on a given hardware platform. This is because the graph-based framework effectively accommodates the backend information of quantum devices. However, more extensive experiments involving tests on different quantum devices and noise models are necessary to thoroughly assess the capability of our GNN to predict the noisy SRE measurements on real quantum devices.

To facilitate benchmarking and support advancements in this field, we release a comprehensive dataset of quantum circuits specifically designed for the nonstabilizerness estimation problem in those three formulations. We highlight that the GNN trained on a specific dataset is not capable of generalize to different circuit structures. However, the GNN trained on the dataset of random quantum circuits provides a useful baseline that allows to significantly reduce the size of the set of circuits to label to achieve a target level of predictive accuracy.

The performance analysis on the experimental results demonstrate that by leveraging the graph-based representation of quantum circuits, the proposed GNN model achieves accurate nonstabilizerness estimation with improved generalization capabilities to larger quantum circuits, in terms of number of qubits, gate count, and entanglement level. We conclude that the proposed GNN model offers an effective and scalable framework for estimating nonstabilizerness across both noiseless and noisy regimes.

\subsection*{Future Directions}
We identify three main future research directions.
The first regards the design of different features that can improve the graph-based quantum circuit representation for the proposed GNN model. The integration of expectation values of local observables, which can be calculated efficiently through the classical shadows protocol~\cite{huang2020predicting}, can be an interesting direction as suggested by previous theoretical~\cite{mello2025retrieving} and experimental results~\cite{lipardi2025study}.

The second future research direction is extending our GNN model to predict different properties of quantum circuits, such as different entropy measures, expectation values over Pauli strings, and relevant observables like Hamiltonians of quantum systems. In this regard, performing a task-agnostic pretraining of the GNN model on our proposed dataset may enable GNN to capture general structural and statistical features of quantum circuits before undergoing problem-dependent finetuning. Pretraining would not only broaden the range of potential applications but also substantially reduce the computational cost associated with training task-specific GNN models.

The third future research direction is the integration of the proposed GNN model within Quantum Architecture Search (QAS) frameworks, to enable the development of SRE-informed QAS techniques~\cite{du2022quantum,zhang2022differentiable}. A particularly suitable framework to embed SRE estimation in quantum architecture search is Monte Carlo Tree Search~\cite{lipardi2025quantum}, where the information regarding the nonstabilizerness can be used in the selection or expansion steps to push the search towards quantum circuits difficult to simulate on a classical device. This direction might be significantly beneficial in the quest for achieving quantum advantage with variational quantum algorithms (VQAs)~\cite{cerezo2021variational}, as it would allow the design of quantum circuits that optimize both the quality of the solution to a target problem and their complexity in terms of classical simulation. Moreover, as evaluating objective functions in QAS requires typically significant computational effort, the proposed GNN can be used as a surrogate model to approximate these objectives. Such integration would enhance QAS efficiency and aligns with the pretraining framework described in the second future direction.

\section*{Data Availability}
The data and code to reproduce the experiments described in the article are publicly available at Zenodo~\cite{lipardi2025dataset} and GitHub: \url{https://github.com/VincenzoLipardi/GNN_SRE}.

\section*{Acknowledgments}
The authors acknowledge the financial support of the Quantum Software Consortium, which facilitated the development of this work and strengthened collaborations through a visitor travel grant. The authors also acknowledge the editor and anonymous reviewers for their constructive feedback, which helped to improve the manuscript.

\section*{Conflict of Interests}
The authors declare no conflict of interests.

\bibliographystyle{unsrturl}
\providecommand{\newblock}{}
\bibliography{bibliography.bib}

\setcounter{section}{0}
\renewcommand{\thesection}{\Alph{section}}
\appsection{Complementary Analysis to the Datasets}{data_generation_appendix}

\appsubsection{Dataset for Stabilizer State Classification}{data_classification_appendix}
The stabilizer state classification problem has been previously investigated in the supervised learning setting~\cite{mello2025retrieving}. However, performing our experiments on exactly the same dataset was not possible for two main reasons. First, the dataset itself is not available. Second, the data generation procedure was stochastic, and without access to the original random seeds or generation metadata, identical reproduction could not be guaranteed. Nevertheless, the original authors provided all the necessary information to replicate and adapt their procedure in our circuit-based problem formulation. 

We generate two datasets, both including all three circuit types. The first is with a fixed number of qubit ($n=18$), which is generated with the same size of the dataset used in~\cite{mello2025retrieving}, in order to have a fair comparison between our GNN and CNN. The second dataset instead enables more general experiments including quantum circuits with number of qubits ranging from $2$ to $25$. 
A schematic summary of the dataset for stabilizer state classification is provided in Table~\ref{tab:dataset_classification}.

\begin{table}[!b]
\centering
\caption{Overview of the dataset for stabilizer state classification.}
\label{tab:dataset_classification}
\renewcommand{\arraystretch}{1.2}
\begin{tabular}{@{}lcccc@{}}
\toprule
\textbf{Dataset} & \textbf{\# Circuits} & \textbf{\# Qubits} & \textbf{Label} & \textbf{SRE} \\ \midrule
\textbf{PS 18} & 15000 & $18$ &  Balanced &  Computed per qubit\\ \midrule
\textbf{CS 18} & 375000 & $18$ & Inherited by PS 18 & Inherited by PS 18\\ \midrule
\textbf{ES 18} & 15000 & $18$ &  Inherited by PS 18 &   - \\\midrule
\textbf{PS 2-25} & 33000 & $2\!-\!25$ &  Balanced &  Computed per qubit \\ \midrule
\textbf{CS 2-25} & 825000 & $2\!-\!25$ &  Inherited by PS 2-25 & Inherited by PS 2-25 \\ \midrule
\textbf{ES 2-25} & 33000 & $2\!-\!25$ &  Inherited by PS 2-25 &  - \\ \bottomrule
\end{tabular}
\end{table}

\begin{figure}[t]
    \centering
    \begin{subfigure}[t]{0.6\textwidth}
        \centering
        \includegraphics[height=6.5cm]{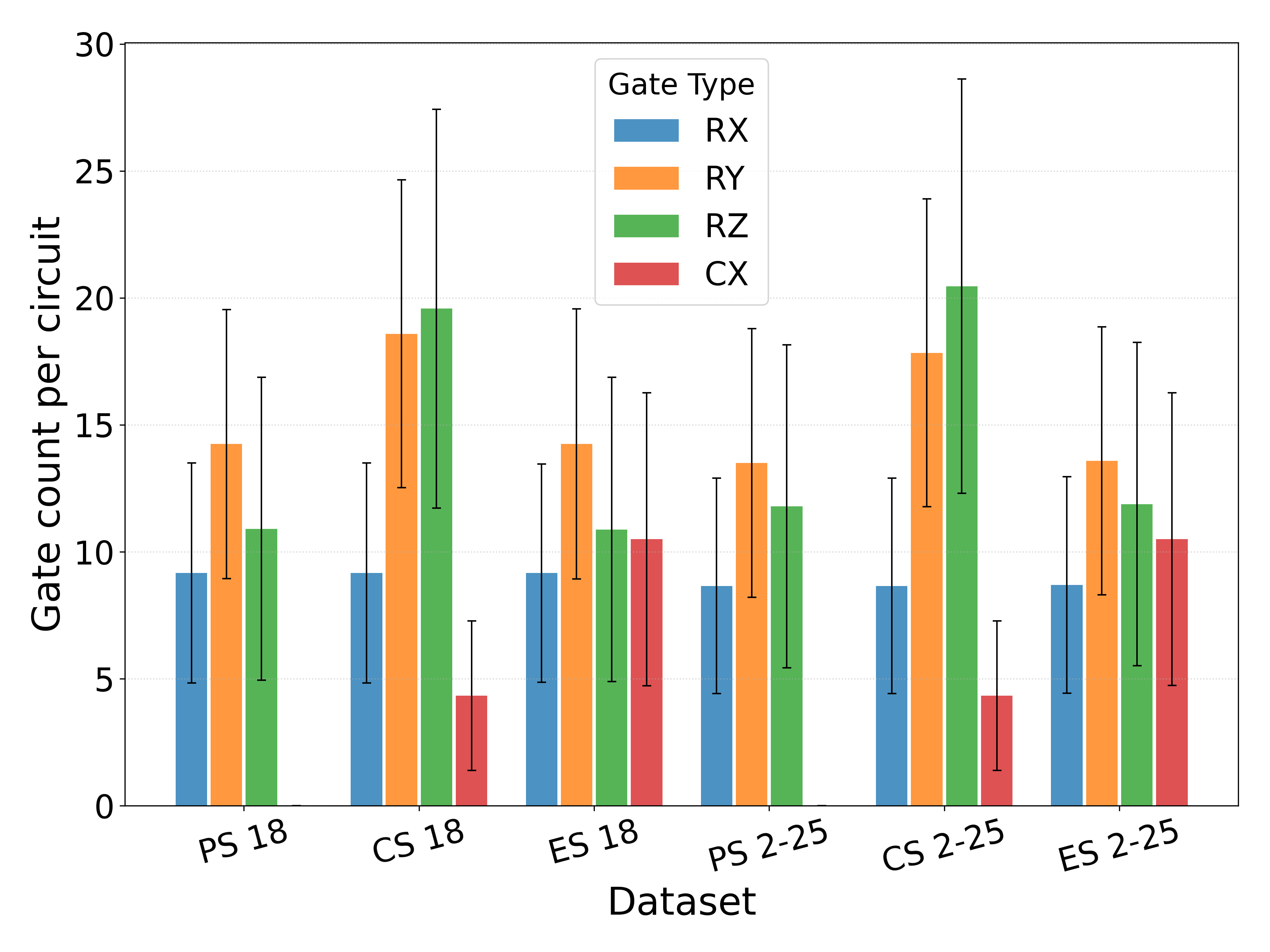} 
        \caption{Gate Count distribution}
        \label{fig:classification_data_a}
    \end{subfigure}
    \begin{subfigure}[t]{0.36\textwidth}
        \centering
        \includegraphics[height=6.4cm]{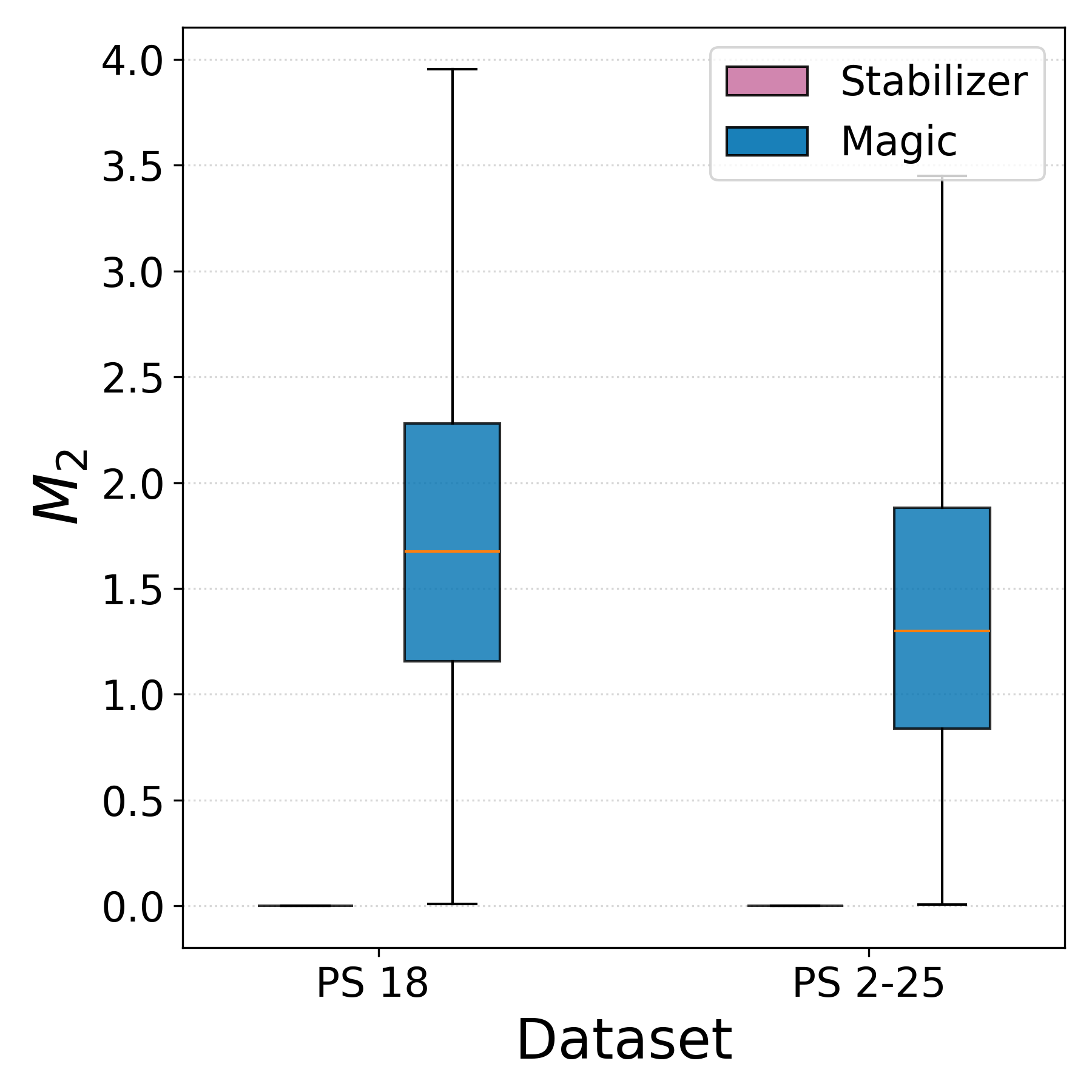} 
        \caption{SRE distribution}
        \label{fig:classification_data_b}
    \end{subfigure}
    \caption{Overview of the quantum circuit composition in the dataset designed for the stabilizer state classification problem and the respective $M_2$ values.}
    \label{fig:data_classification}
\end{figure}

Each circuit is assigned a total number of gates $G$, uniformly sampled in $[10, 100]$.
Stabilizer states are generated by applying gates randomly selected from the set of Clifford single-qubit rotations, while magic states involve single-qubit rotation gates with both Clifford angles and random angles, which inject magic into the circuits. For stabilizer states, each gate is chosen randomly among the set of allowed Clifford gates: \[
\mathcal{S}_S = 
\Big\{
\texttt{RX}(\theta), \texttt{RY}(\theta), \texttt{RZ}(\theta)
\mid 
\theta \in \{\pm \frac{\pi}{2}, \pm \pi, \pm \frac{3\pi}{2}\}
\Big\}
\cup
\{\texttt{H}, \texttt{S}\},
\]
where the Hadamard gate and Phase gate are expressed in terms of rotation gates:\\
\texttt{H} = \texttt{RY}$(\frac{\pi}{2})$\texttt{RZ}$(\pi)$ and \texttt{S} = \texttt{RZ}$(\frac{\pi}{2})$.
For magic states, each gate is chosen from an extended gate set that includes both Clifford gates and single-qubit rotations with random angles:
\[
\mathcal{S}_M = 
\mathcal{S}_S 
\cup
\{\texttt{RX}(\theta), \texttt{RY}(\theta), \texttt{RZ}(\theta) 
\mid \theta \in (0, 2\pi)\}.
\]
Each magic states is assigned a ratio $r_M$ of magic gates over the total number of gates uniformly sampled between $0.3$ and $1$, so that each magic circuit contains approximately $r_M \cdot G$ gates with random angles, effectively injecting varying degrees of nonstabilizerness into the circuit.

The presence of CS circuits in the dataset allows to investigate the generalization capabilities of the models to unseen entangled states, while keeping the same labels and $M_2$ values, in line with the procedure used in~\cite{mello2025retrieving}. The Entangled States (ES) subset, on the other hand, enables a further step in this analysis by testing model performance on circuits characterized by substantially different structural patterns and, in general, a higher degree of entanglement.
Figure~\ref{fig:data_classification} shows the heterogeneous quantum circuit dataset generated with our procedure, spanning a broad range of qubit number, circuit depth, circuit structure, entanglement and $M_2$. Figure~\ref{fig:classification_data_a} illustrates the diversity of quantum circuits across the three different types in terms of gate counts and $M_2$. It is important to note that our procedure generates also quantum circuits with significantly low $M_2$ values. For instance, in the case of $18$-qubit circuits, the minimum $M_2$ value obtained in our dataset is $0.01$, compared to $1.80$ reported in~\cite{mello2025retrieving}. Moreover, in the dataset PS 2-25, we observe a minimum $M_2$ value of $0.006$, see Figure~\ref{fig:classification_data_b}. This is relevant, as the performance of machine learning models on the stabilizer state classification problem is typically correlated with the density of $M_2$ in the number of qubits $n$: $m_2=\frac{M_2}{n}$. Models tend to achieve higher accuracy for states with higher $m_2$, because of the increased separability of the magic states from the stabilizer states. This effect is also shown in the experimental results discussed in~\cite{mello2025retrieving} and further confirmed by our own results obtained using graph neural networks (GNNs).

From a geometric perspective, stabilizer states form a discrete subset of the Hilbert space, i.e., the states reachable via the Clifford group acting on the computational basis. On the other hand, magic states lie outside this subset, covering regions of the Hilbert space not accessible through Clifford operations alone. The stochastic mixing of Clifford and non-Clifford gates thus corresponds to sampling points with varying distances from the set of stabilizer states, parameterizing to a certain extent the degree of magic in the resulting quantum states.

\begin{figure}[hb]
    \centering
    \includegraphics[width=0.6\linewidth]{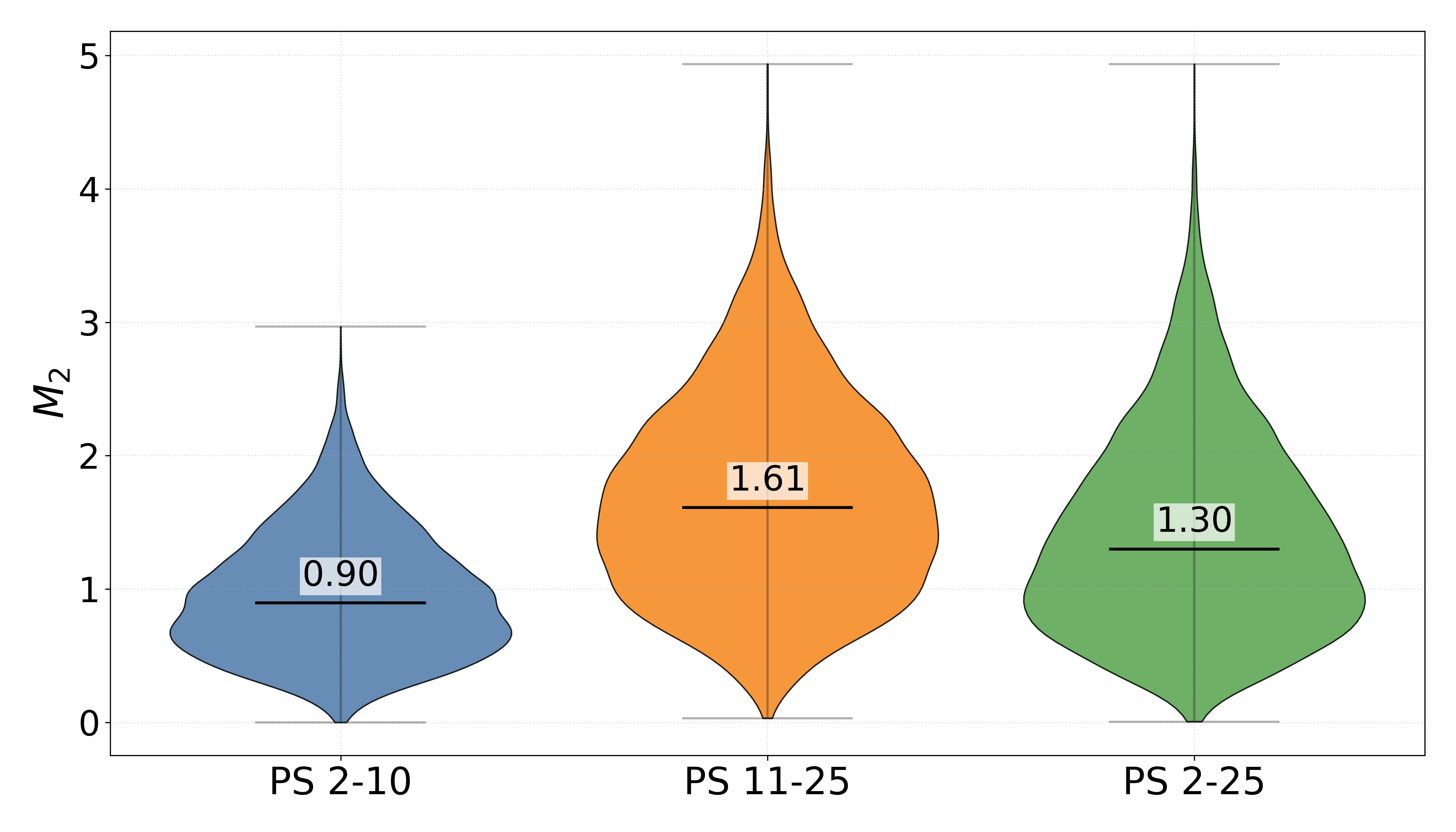}
    \caption{$M_2$ value distribution in datasets of product states.}
    \label{fig:sre_based_data_distribution}
\end{figure}

\appsubsection{Dataset for SRE-based Classification}{data_classification_sre_appendix}
Here, we provide a further analysis of the dataset generated for the SRE-based classification problem, which we recall it results from a simple procedure applied to the dataset generated for the stabilizer state classification problem. Specifically, the stabilizer states (labeled as $0$) are removed from the dataset, and among the remaining circuits (originally labeled as $1$), those with $M_2$ below a given threshold are relabeled as $0$.
The violin plots in Figure~\ref{fig:sre_based_data_distribution} illustrates the distribution of $M_2$ values across the product-state datasets. The horizontal lines in the plot represent the median $M_2$ value for each dataset shown along the x-axis. As expected, quantum circuits with qubit numbers between $2$ and $10$ exhibit a narrower range of $M_2$ values compared to circuits with qubit numbers between $11$ and $25$. 

We note that by choosing the median as the threshold $M_2$ value in our dataset we ensure that is balanced between the two labels: low-$M_2$ and high-$M_2$. However, the threshold choice is a way to parametrizes the difficulty of the classification task. In particular, for low $M_2$ threshold the SRE-based classification becomes a physics-motivated problem where the goal is to distinguish between stabilizer and magic states on a real quantum hardware. In this case, the threshold will represent an indication of the quantum noise impact on the $M_2$ value of the stabilizer states.

\begin{figure}[!t]
   \centering

    \begin{subfigure}[t]{0.49\textwidth}
        \centering
        \includegraphics[width=\linewidth]{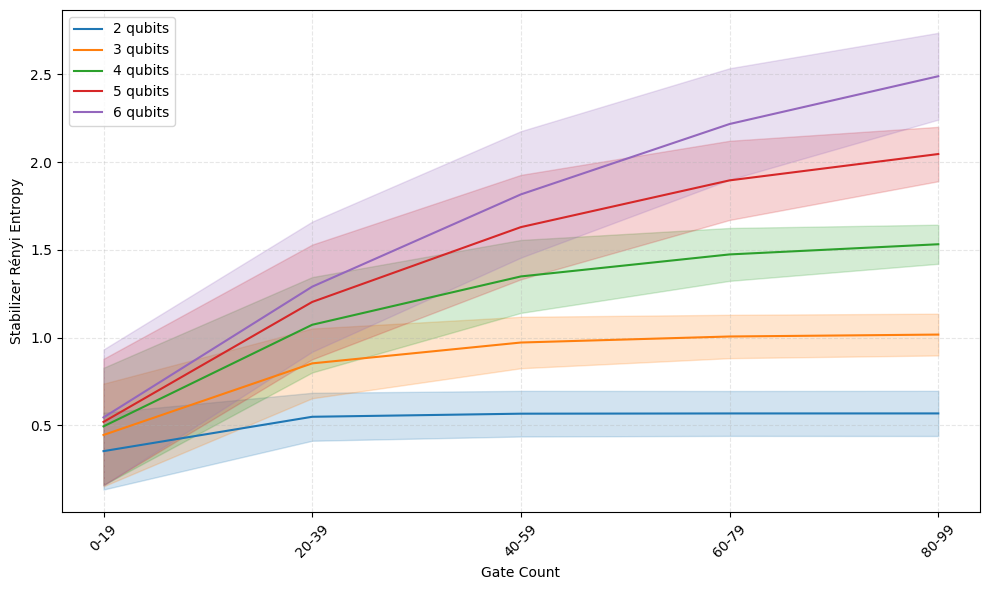}
        \caption{RQC}
        \label{fig:random_d}
    \end{subfigure}
    \hfill
    \begin{subfigure}[t]{0.49\textwidth}
        \centering
        \includegraphics[width=\linewidth]{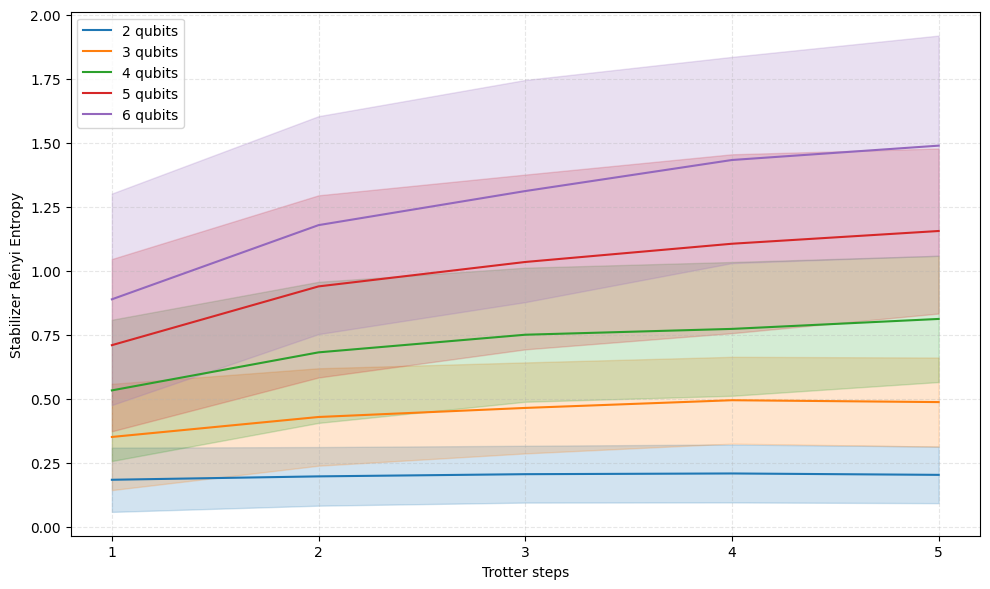}
        \caption{TIM }
        \label{fig:tim_d}
    \end{subfigure}
    \caption{Overview of the distribution of the $M_2$ values across the quantum circuits in the dataset designed for the SRE estimation problem.  }
    \label{fig:data}
\end{figure}

\begin{figure}[!t]
    \centering
    \begin{subfigure}[t]{0.49\textwidth}
        \centering
        \includegraphics[width=\linewidth]{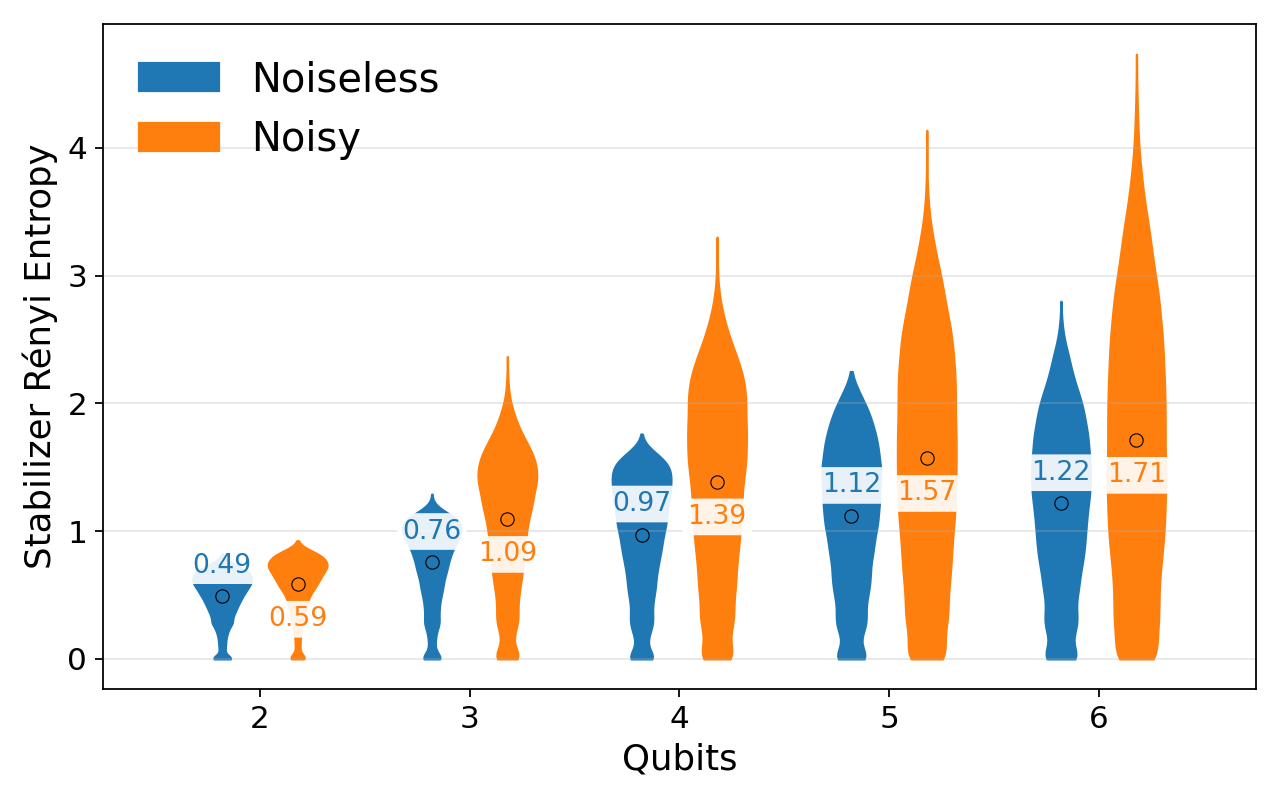}
        \caption{RQC Noisy}
        \label{fig:random_noise}
    \end{subfigure}
    \begin{subfigure}[t]{0.49\textwidth}
        \centering
        \includegraphics[width=\linewidth]{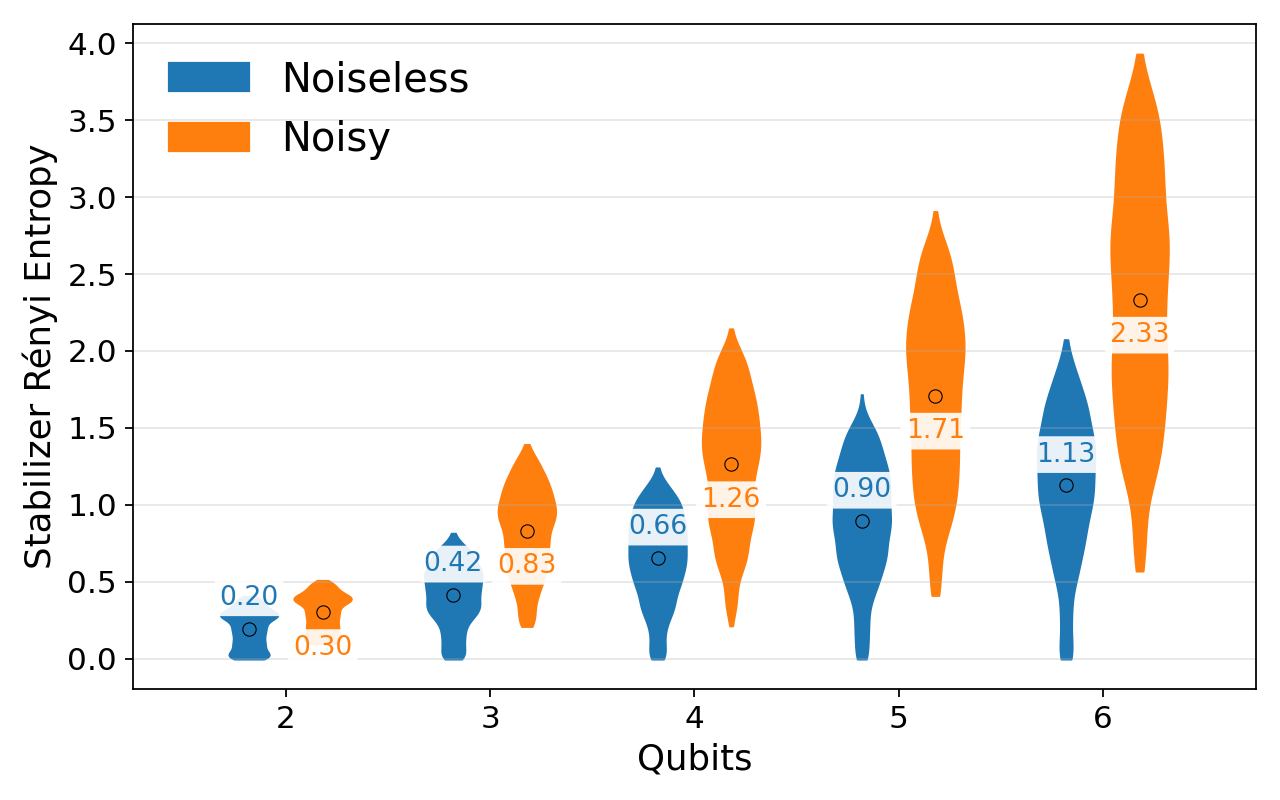}
        \caption{TIM Noisy}
        \label{fig:tim_noise}
    \end{subfigure}
    \caption{SRE value distributions of quantum circuits grouped by qubit number in the RQC and TIM noisy datasets are shown in Figure~\ref{fig:random_noise} and~\ref{fig:tim_noise}, respectively. The $\tilde{M_2}$} values are computed under simulated noise using the IBM \texttt{FakeOslo} backend.
    \label{fig:data_noise}
\end{figure}

\appsubsection{Dataset for Stabilizer R\'enyi Entropy Estimation}{data_regression_appendix}
 In this section we further analyze the dataset generated for the SRE estimation problem, introduced in Section~\ref{data_regression}.
Figures~\ref{fig:random_d} and~\ref{fig:tim_d} presents the average $M_2$ values, together with their standard deviations (shaded regions), for quantum circuits grouped by qubit number as a function of circuit depth. On the other hand, Figure~\ref{fig:data_noise} shows the corresponding distributions obtained from noisy simulations in comparison to the noiseless case. Figures~\ref{fig:random_noise} and~\ref{fig:tim_noise} demonstrate that noisy quantum simulations have a stronger impact on the TIM dataset than on the RQC dataset: on average, the $M_2$ values in the TIM dataset shift more towards higher values, and the overall distributions become more stretched.

Table~\ref{tab:dataset_regression} provides a schematic overview of the datasets used for the SRE estimation problem.
Figure~\ref{fig:data} illustrates the $M_2$ value distribution under noiseless simulations for the RQC and TIM datasets.

\begin{table}[!b]
\centering
\caption{Overview of the dataset for the SRE estimation problem.}
\label{tab:dataset_regression}
\renewcommand{\arraystretch}{1.2}
\begin{tabular}{@{}lcccc@{}}
\toprule
\textbf{Dataset} & \textbf{\# Circuits} & \textbf{\# Qubits} & \textbf{Depth} & \textbf{Extrapolation} \\ \midrule
\textbf{RQC} & 250000 & $2\!-\!6$ & $1\!-\!100$ & Qubits and Gate count \\ \midrule
\textbf{RQC Noisy} & 100000 & $2\!-\!6$ & $1\!-\!60$ & Qubits and Gate count \\ \midrule
\textbf{TIM} & 25000 & $2\!-\!6$ & $1\!-\!5$ & Qubits and Trotter steps \\ \bottomrule
\textbf{TIM Noisy} & 15000 & $2\!-\!6$ & $1\!-\!3$ & Qubits and Trotter steps  \\ \bottomrule
\end{tabular}
\end{table}

In the RQC dataset, once we fix the number of qubits \(n\), the circuits are generated by sampling the number of gates uniformly from the range \(G=[0, 100]\). Starting with an empty quantum circuit, gates are sequentially added, each randomly picked from a universal gate set comprising the \texttt{CNOT} gate and the three single-qubit rotation gates: \texttt{RX}, \texttt{RY}, \texttt{RZ}.  
In the TIM dataset, each quantum circuit represents a discrete-time approximation of the transverse-field Ising hamiltonian on a one-dimensional chain with nearest-neighbor interactions:
\[
H = -J \sum_{i=1}^{n-1} Z_i Z_{i+1} - h \sum_{i=1}^{n} X_i,
\]
where \(Z_i\) and \(X_i\) are Pauli operators acting on qubit \(i\), \(J\) denotes the interaction strength, and \(h\) the transverse-field strength.  
The time-evolution operator \(U(t) = e^{-i H t}\) is simulated using a first-order Trotter--Suzuki decomposition~\cite{nielsen2010quantum}. Each Trotter step is constructed as a sequence of quantum gates: the two-qubit interaction \(e^{-i \theta Z_i Z_{i+1}}\), with \(\theta = J \Delta t\), is realized by the standard \(\texttt{CNOT}_{i,i+1}\)–\(\texttt{RZ}(2\theta)_{i+1}\)–\(\texttt{CNOT}_{i,i+1}\) construction, while the transverse-field term \(e^{-i \phi X_i}\), with \(\phi = h \Delta t\), is implemented as a single-qubit rotation \(\texttt{RX}(2\phi)_i\). The circuits span various values of the angle parameters \(\theta\), \(\phi\), with the number of Trotter steps $T$ ranging from one to five. 

\appsection{Complementary Analysis to the Experimental Results}{complementary_analysis}
In this section, we provide additional experiments and analysis to Section~\ref{results}.

\begin{figure}[!t]
    \centering
    \begin{subfigure}[t]{0.49\textwidth}
        \centering
        \includegraphics[width=\linewidth]{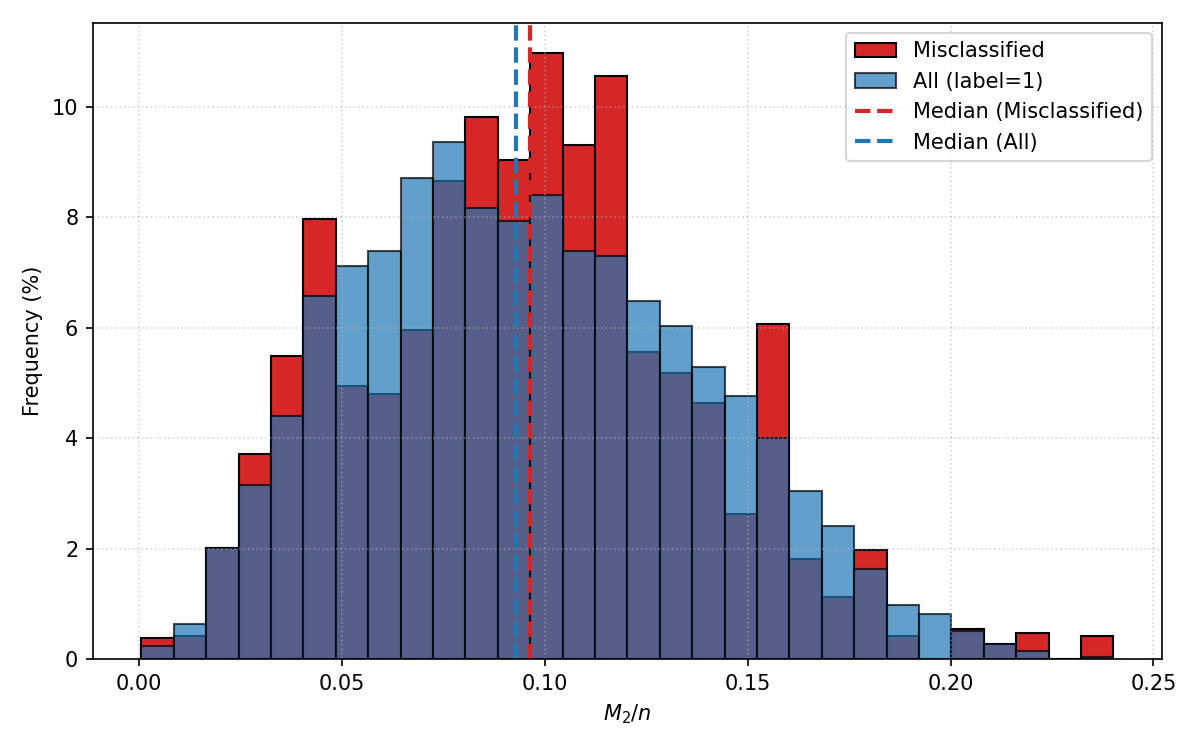}
        \caption{Trained on PS 18}
        \label{fig:misclassified_distribution}
    \end{subfigure}
    \begin{subfigure}[t]{0.49\textwidth}
        \centering
        \includegraphics[width=\linewidth]{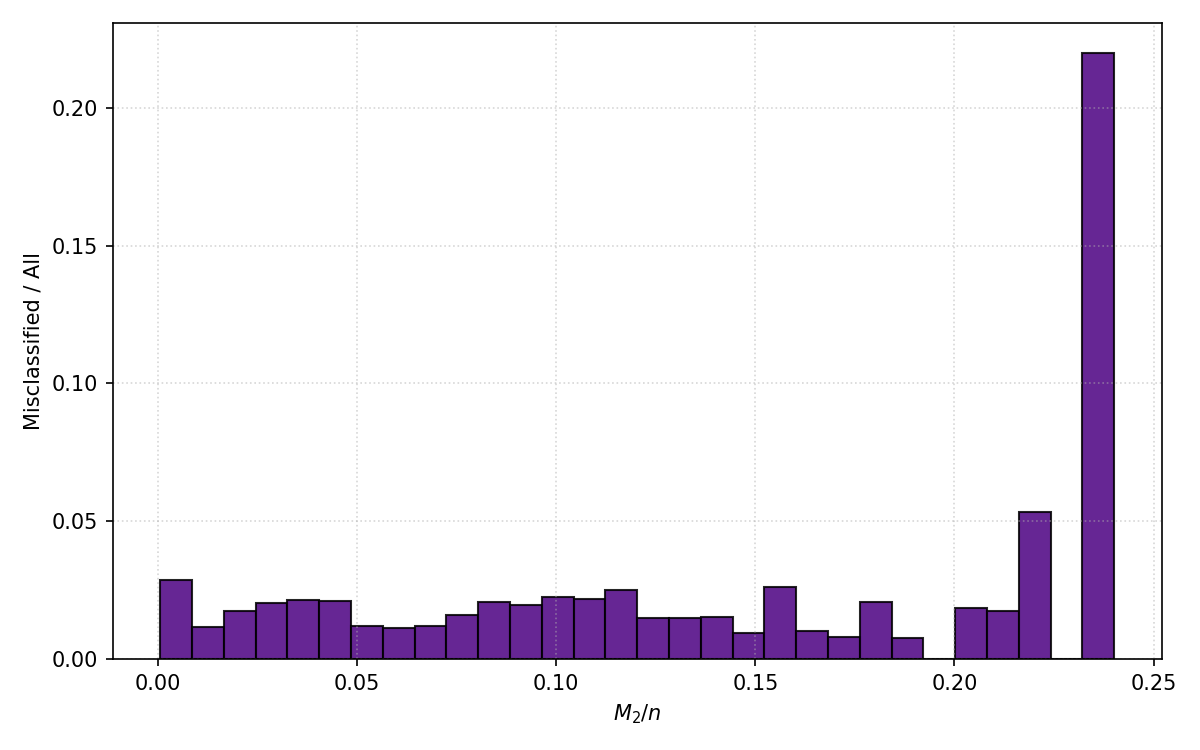}
        \caption{Trained on PS 2-10}
        \label{fig:misclassified_all}
    \end{subfigure}
    \caption{Evaluation of the GNN model trained on PS 18 over the circuits in CS 18 for different SRE density values ($M_2/n$).}
    \label{fig:result_misclassified}
\end{figure}

\appsubsection{Performance Analysis for the Stabilizer State Classification}{ssc_complementary}
In Section~\ref{data_classification} we discussed the importance of including quantum circuits with low $M_2$ in the datasets the classification task is easier for magic states with high $M_2$ density: $m_2=M_2/n$.
Figure~\ref{fig:result_misclassified} presents the results of the GNN model trained on PS 18 and evaluated on CS 18 for different values of $m_2$, where we divided the range from $0$ to $0.24$ in $30$ bins. Specifically, Figure~\ref{fig:misclassified_distribution} plots in red the data distribution at different ranges of $m_2$ compared with the distribution of the misclassified circuits from CS 18 in blue. The vertical lines represent the medians of the two distributions, which are $0.093$ and $0.096$, respectively. Figure~\ref{fig:misclassified_all} reports the ratio of misclassified circuits to the total number of circuits in CS 18 for each of the $30$ bins. The peak visible at the far right of the plot can be attributed to the few circuits located in the last bin, which correspond to the tail of the distribution shown in Figure~\ref{fig:misclassified_distribution}. In fact, in the last bin there are $11$ circuits that are misclassified out of only $50$ total circuits. Note that in both plots of Figure~\ref{fig:result_misclassified}, we only consider the circuits labeled as magic states, which account for half of the CS 18 dataset ($187500$ circuits), since for all other circuits $m_2=0$. Overall, these experimental results indicate that the GNN classification performance remains robust across different ranges of $m_2$ values.

\begin{figure}[!ht]
    \centering
    \includegraphics[width=0.60\linewidth]{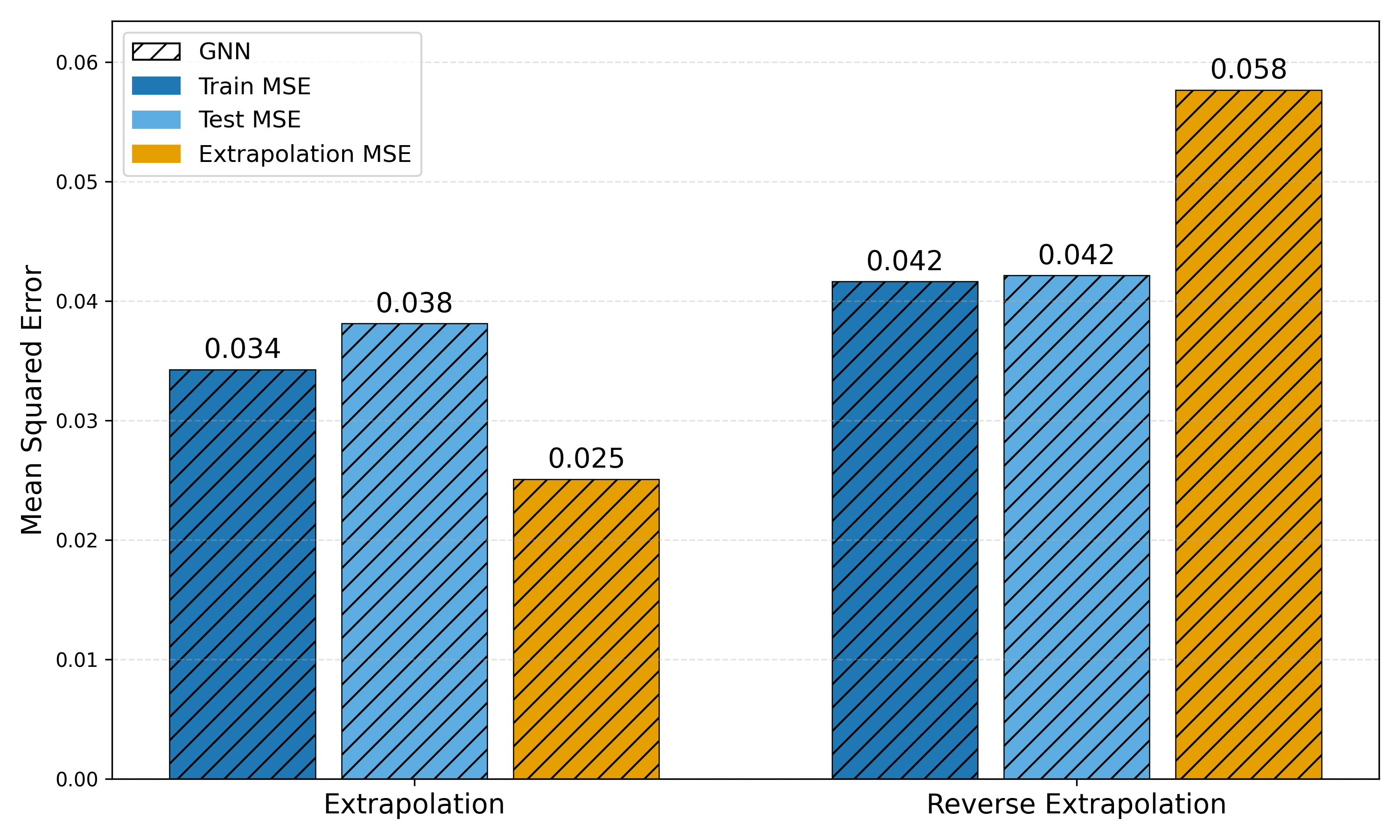}
    \caption{Extrapolation performance of the GNN model with respect to gate counts in the RQC dataset. On the left, the bars correspond to the canonical extrapolation on larger circuits, while on the right, they correspond to the extrapolation on smaller circuits.}
    \label{fig:reverse_extrapolation}
\end{figure}

\appsubsection{Performance Analysis for the Stabilizer Rényi Entropy Estimation}{sre_complementary}
It is important to focus on the extrapolation MSE obtained by the proposed GNN on the RQC dataset when extrapolating on gate counts, as it is unexpectedly lower than the training MSE. 
Figure~\ref{fig:random_d} shows that the SRE distribution of the quantum circuits in RQC becomes narrower, with the standard deviation decreasing as the gate count increases. Since $M_2$ is bounded above by $SRE_{max}(n)=\ln (\frac{2^n+1}{2})$~\cite{leone2022stabilizer}, for a quantum system of $n$ qubits, higher gate counts that include non-Clifford gates drive the system towards a more stable regime with smaller variance. To further investigate this effect, we perform an additional experiment in which we train our GNN on quantum circuits with the largest gate counts, namely from $20$ to $99$, and extrapolate on the shallower circuits with gate counts from $1$ to $19$. In this setup, we require the model to extrapolate on a subset of circuits with a wider interval of $M_2$, higher standard deviation, and less separation. As a result, extrapolating in the new setup should be more difficult. Figure~\ref{fig:reverse_extrapolation} presents a direct comparison between two extrapolation experiments. The bars on the left, relate to the GNN extrapolation on higher gate counts. On the right, there are the bars related to the extrapolation to unseen circuits with lower gate counts. As expected, the extrapolation MSE in the latter setting is higher than in the former, and also exceeds the training MSE.

\appsubsection{Extended Analysis to the Extrapolation in Qubits}{qubit_extrapolation_appendix}
In this section, we extend our experiments to evaluate the generalization performance of the GNN on circuits with higher qubit counts, beyond those tested in Section~\ref{exp_regression} and in the previous work~\cite{lipardi2025study}. The limitations on the number of circuits arise from the computational time required to compute $M_2$ for completely unstructured circuits in the RQC dataset, which increases exponentially with the number of qubits. Nevertheless, we expanded the RQC dataset to include circuits with up to $6$ qubits, adding $200$ additional circuits equally divided between $7$- and $8$-qubit circuits. 

\begin{figure}[!ht]
    \centering
    \includegraphics[width=0.5\linewidth]{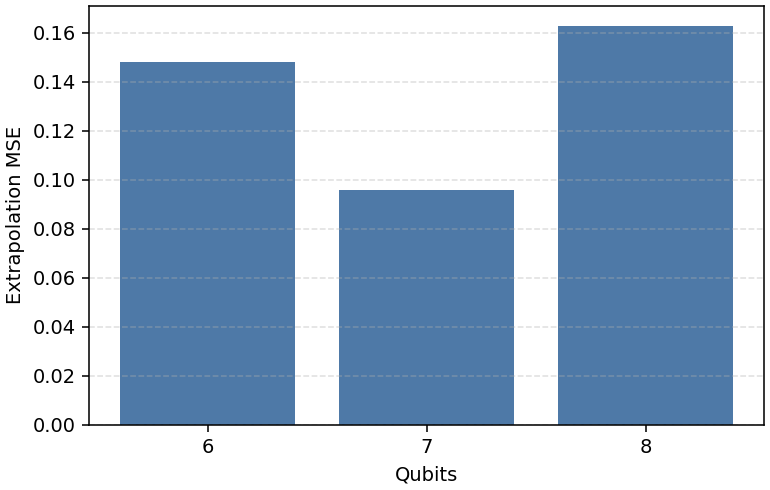}
    \caption{Extrapolation Mean Squared Error (MSE) in the number of qubits for the GNN.}
    \label{fig:extrapolation_qubit_app}
\end{figure}

The GNN described in Section~\ref{exp_regression} is trained on circuits from the RQC dataset containing up to $5$ qubits. The extrapolation MSE reported in Figure~\ref{gnn_extrapolation_qubits} is computed on the remaining $6$-qubit circuits, which are $50000$. Figure~\ref{fig:extrapolation_qubit_app} recalls the extrapolation MSE achieved on the $6$-qubit circuits and extends the analysis to the newly added $7$- and $8$-qubit circuits.

We observe that the extrapolation MSE achieved on the few additional circuits is consistent with that obtained on the more statistically significant set of $6$-qubit circuits. The reduced extrapolation MSE observed for the $7$-qubit circuits can likely be attributed to a favorable random generation of circuits. In general, the extrapolation MSE increases for circuits with a number of qubits farther from the training set, although it remains reasonably stable overall. This observation also aligns with the experiments on generalization performance with respect to the number of qubits in the classification setting.

\appsection{Generalization to Structured Quantum Circuits}{generalization_to_structured_circuits}
In this section, we study the generalization capabilities of the proposed GNN approach when trained on the unstructured RQC dataset and tested on two different structured classes of quantum circuits. The first is derived by the one-dimensional TIM model and the second one-dimensional Heisenberg model. Similarly to the TIM model, which we extensively discussed in Appendix~\ref{data_regression_appendix}, in the Heisenberg model, each quantum circuit represents a discrete-time approximation of the continuous-time evolution generated by the Heisenberg Hamiltonian. For a one-dimensional chain of $n$ qubits with nearest-neighbor interactions, the Hamiltonian is given by:
\begin{equation}
    H_{Heis} = \sum_{i=1}^{n-1} J_X X_iX_j+J_Y Y_iY_j+J_Z Z_iZ_j
\end{equation}
where $J_X$, $J_Y$, and $J_Z$ are real coupling constants.

The corresponding time-evolution operator $U(t) = e^{-i t H_{\mathrm{Heis}}}$ is approximated using a first-order Trotter--Suzuki decomposition~\cite{nielsen2010quantum}:
\begin{equation}
    U(t)
    \approx
    \left[
        \prod_{i=1}^{n-1}
        e^{-i \theta_X X_i X_{i+1}}
        e^{-i \theta_Y Y_i Y_{i+1}}
        e^{-i \theta_Z Z_i Z_{i+1}}
    \right]^T,
\end{equation}
where $\theta_\alpha = J_\alpha \Delta t$ for $\alpha \in \{X,Y,Z\}$.
Each two-qubit interaction term is implemented using elementary quantum gates. The $ZZ$ interaction $e^{-i \theta_Z Z_i Z_{i+1}}$ is realized by the gate sequence $\texttt{CNOT}_{i,i+1}$–$\texttt{RZ}(2\theta_Z)_{i+1}$–$\texttt{CNOT}_{i,i+1}$.
The $XX$ interaction $e^{-i \theta_X X_i X_{i+1}}$ is obtained by conjugating the $ZZ$ interaction with Hadamard gates on both qubits, while the $YY$ interaction $e^{-i \theta_Y Y_i Y_{i+1}}$ is implemented by conjugating the $ZZ$ interaction with $R_X(\pi/2)$ and $R_X(-\pi/2)$ rotations. The circuits span various values of the angle parameters $\theta_{\alpha}$, with the number of Trotter steps $T$ ranging from one to five. 

\begin{figure}[!ht]
    \centering
    \includegraphics[width=0.75\linewidth]{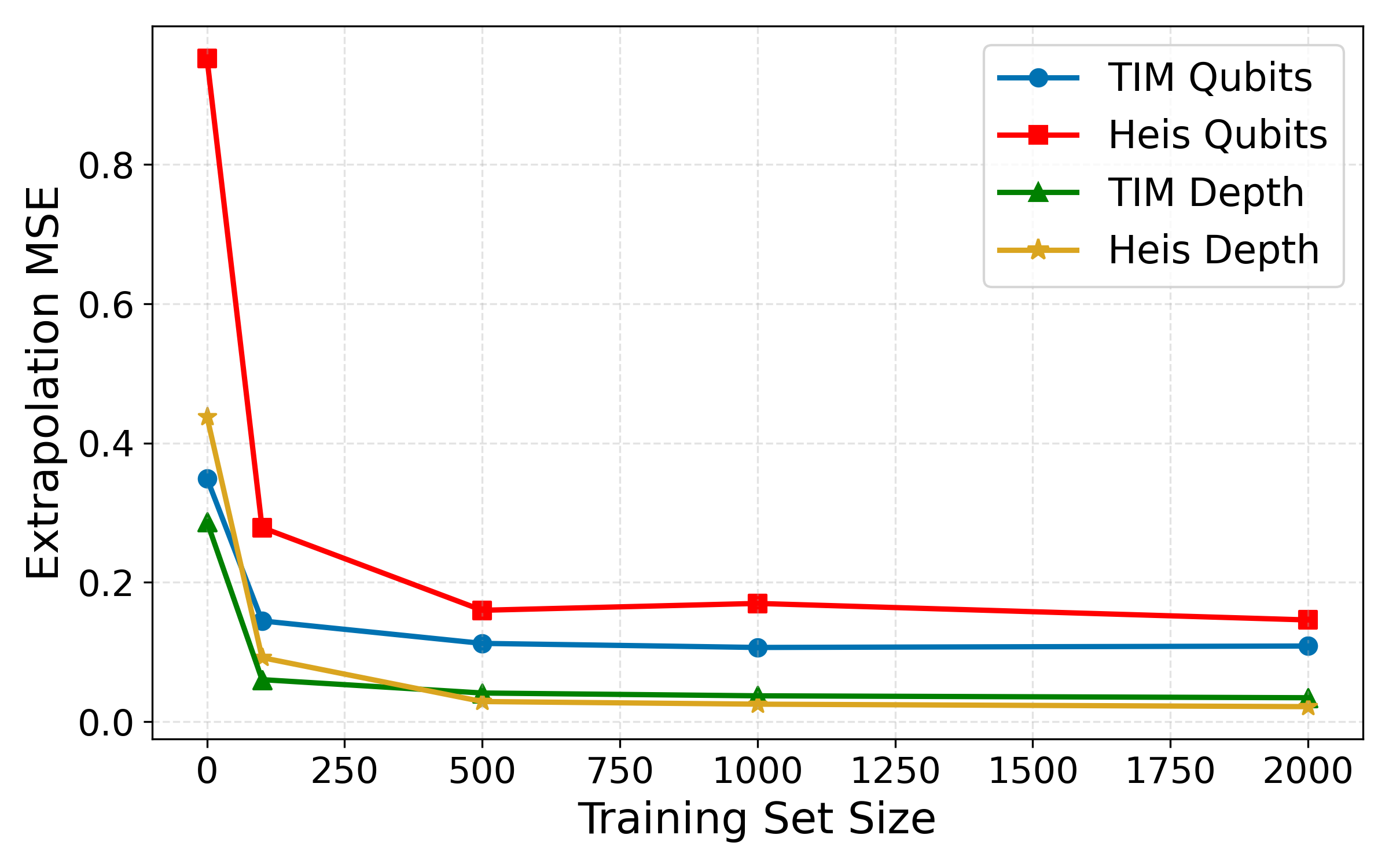}
    \caption{Extrapolation MSE on two class of structured quantum circuits obtained by the GNN trained on the unstructured RQC dataset. }
    \label{fig:finetuning}
\end{figure}

The experimental results indicate that when the GNN is trained on the unstructured RQC dataset, the predictions do not generalize well to structured circuits. However, by fine-tuning the model on a small number of circuits from the corresponding class of quantum circuits, the mean squared error (MSE) converges to values comparable to those reported in Section~\ref{exp_regression}. Figure~\ref{fig:finetuning} illustrates the extrapolation MSE on both structured datasets under the configuration in which testing is performed on unseen numbers of qubits and gate counts. We observe that fine-tuning on as few as $50$ circuits leads to a substantial decrease in MSE, and after $500$ circuits, the MSE approaches the values reported in Figure~\ref{gnn_extrapolation_qubits} and Figure~\ref{gnn_extrapolation_depth}, where the training set includes $16000$ circuits.

\begin{revision}
\appsection{Benchmark against Analytic Estimations}{appendix_analytic}
In this section, we compare the $M_2$ estimates obtained by the proposed GNN to those derived from an analytical formula for the stabilizer linear entropy presented in~\cite{leone2022stabilizer}.

The stabilizer linear entropy $M_{\mathrm{lin}}$ of a quantum state $\rho$ is defined as
\[
M_{\mathrm{lin}}(\rho) := 1 - d \left\| \Xi(\rho) \right\|_2^2,
\]
where $d=2^n$ and $\Xi$ is the $4^n$-dimensional vector whose entries are the normalized squared expectation values of the $4^n$ Pauli strings $P$, denoted by $\Xi_P(\ket{\psi})$, also defined in Section~\ref{sre}. 
The quantities $M_2$ and $M_{\mathrm{lin}}$ satisfy the inequality~\cite{leone2022stabilizer}:
 $M_2 \geq -\ln\!\left(1 - M_{\mathrm{lin}}\right)$.
For $k$-doped random quantum circuits, the monotonicity of the linear non-stabilizing power allows an analytic estimate of the expectation value
\begin{equation}
\mathbb{E}_{C_k}\!\left[M_{\mathrm{lin}}(U)\right]=1 - \frac{1}{3 + d}\left(4 + (d - 1)\, f(\theta)^k\right),
\label{eq_13}
\end{equation}
where $\theta$ is the angle of the Phase gate introducing magic in the circuit and 
\[
f(\theta)
=
\frac{
7d^2 - 3d + d(d+3)\cos(4\theta) - 8
}{
8(d^2 - 1)
}
\]
satisfies $f(\theta)\le 1$, with equality for Clifford gates.
Figure~\ref{fig:ftheta} shows $f(\theta)$ for representative values of $d$.
\begin{figure}[!ht]
    \centering
    \begin{subfigure}[t]{0.49\textwidth}
        \centering
        \includegraphics[width=\linewidth]{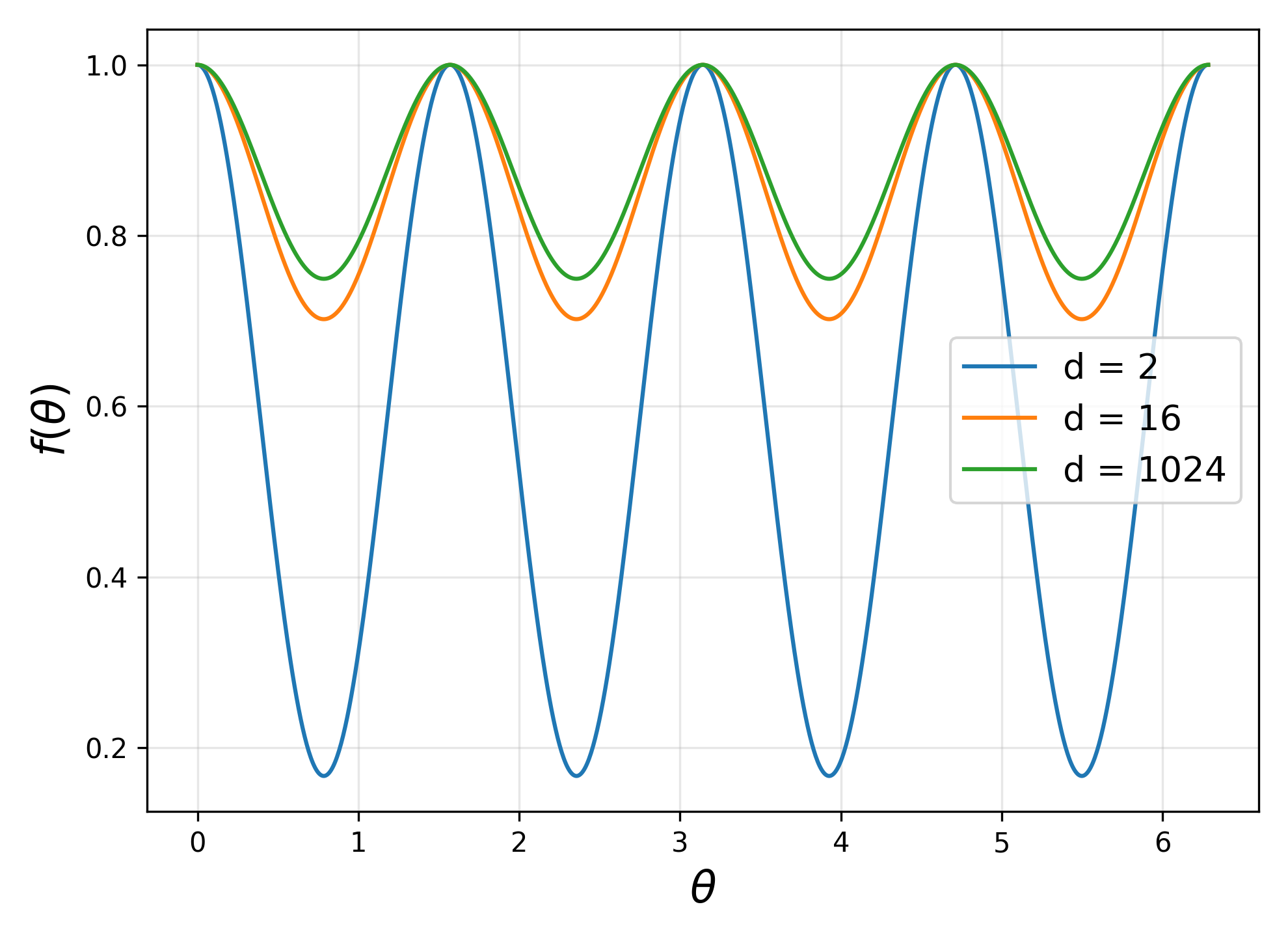}
        \caption{$f(\theta)$ function from Equation~\ref{eq_13}}
        \label{fig:ftheta}
    \end{subfigure}
    \begin{subfigure}[t]{0.49\textwidth}
        \centering
        \includegraphics[width=\linewidth]{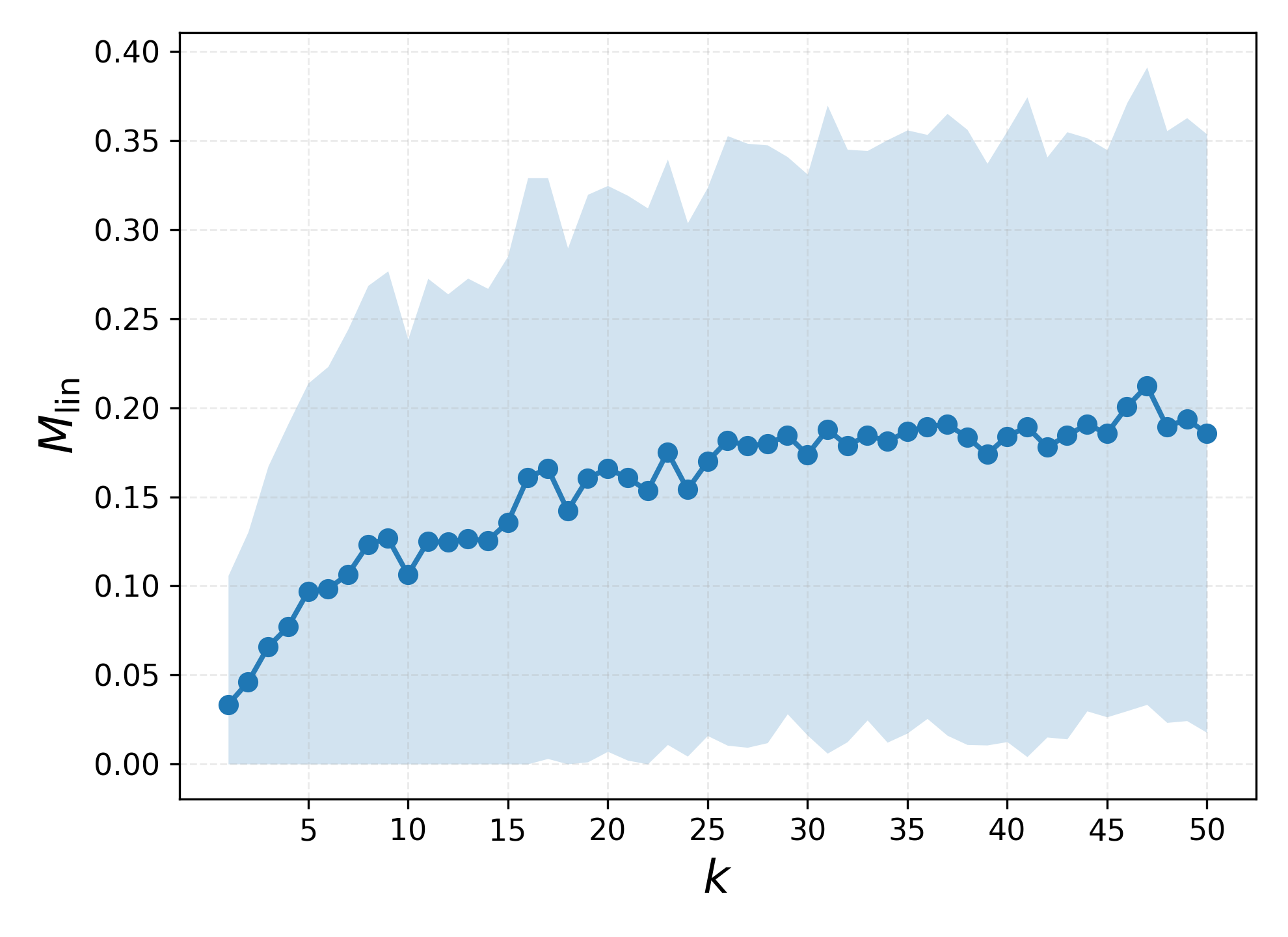}
        \caption{$M_{lin}$ value distribution}
        \label{fig:data_analytic}
    \end{subfigure}
    \caption{$k$-doped random quantum circuits generated to compare the GNN prediction with the analytic estimation.}
    \label{fig:dataset_analytic}
\end{figure}

To benchmark the GNN against this analytic prediction, we generate a benchmark dataset of \(4\)-qubit circuits. We first sampled 100 random Clifford base circuits (single-qubit Clifford rotations and CNOT gates). For each base circuit, we generate a family of doped variants by inserting \(k\) non-Clifford phase operations, with \(k\in\{1,\dots,50\}\). This results in \(100\) circuits per \(k\) and \(5000\) circuits in total. Non-Clifford doping is implemented exclusively as \(R_z(\theta)\) gates (equivalent to \(P(\theta)\) up to a global phase), placed in random qubits and position. Each base circuit is assigned a fixed random non-Clifford angle \(\theta\) (uniformly sampled in $[0,2\pi)$), shared across all its \(k\)-variants to analyze the performance in terms of the doping depth. Each circuit was labeled with its $M_2$ value, and $M_{\mathrm{lin}}$ was obtained from $M_{\mathrm{lin}} = 1 - e^{-M_2}$. The resulting $M_{\mathrm{lin}}$ distribution as a function of the doping depth $k$ is shown in Figure~\ref{fig:data_analytic}.

On this set of $k$-doped circuits, we benchmark the GNN (trained on the general RQC dataset) against the analytic estimate derived from Eq.~\ref{eq_13}. Figure~\ref{fig:results_analytic} reports the comparison in terms of the mean squared error (MSE) with respect to the true $M_{\mathrm{lin}}$.
The results indicate that the GNN achieves lower MSE than the analytic prediction. As shown in Fig.~\ref{fig:k_analytic}, the MSE of the GNN estimate (red line) consistently lies below that of the analytic estimate (blue line). The improvement can be attributed to the more informative input representation exploited by the GNN. Specifically, the GNN operates on a graph-based representation encoding the full circuit structure, including the angles and positions of all gates. In contrast, the analytic formula depends only on the dimension of the Hilbert space $d=2^n$, the number of non-Clifford gates $k$, and the angle $\theta$ of the Phase gate that injects magic into the circuit. Figure~\ref{fig:theta_analytic} illustrates the performance in terms of MSE as a function of $\theta$. The dataset contains $100$ distinct values of $\theta$, for better readability, the figure reports the averaged trend. Although the GNN generally outperforms the analytic estimate, the latter achieves lower error near the maxima of the function $f(\theta)$ (corresponding to Clifford gates), where it locally provides more accurate predictions than the GNN.
\begin{figure}[!hb]
    \centering
    \begin{subfigure}[t]{0.49\textwidth}
        \centering
        \includegraphics[width=\linewidth]{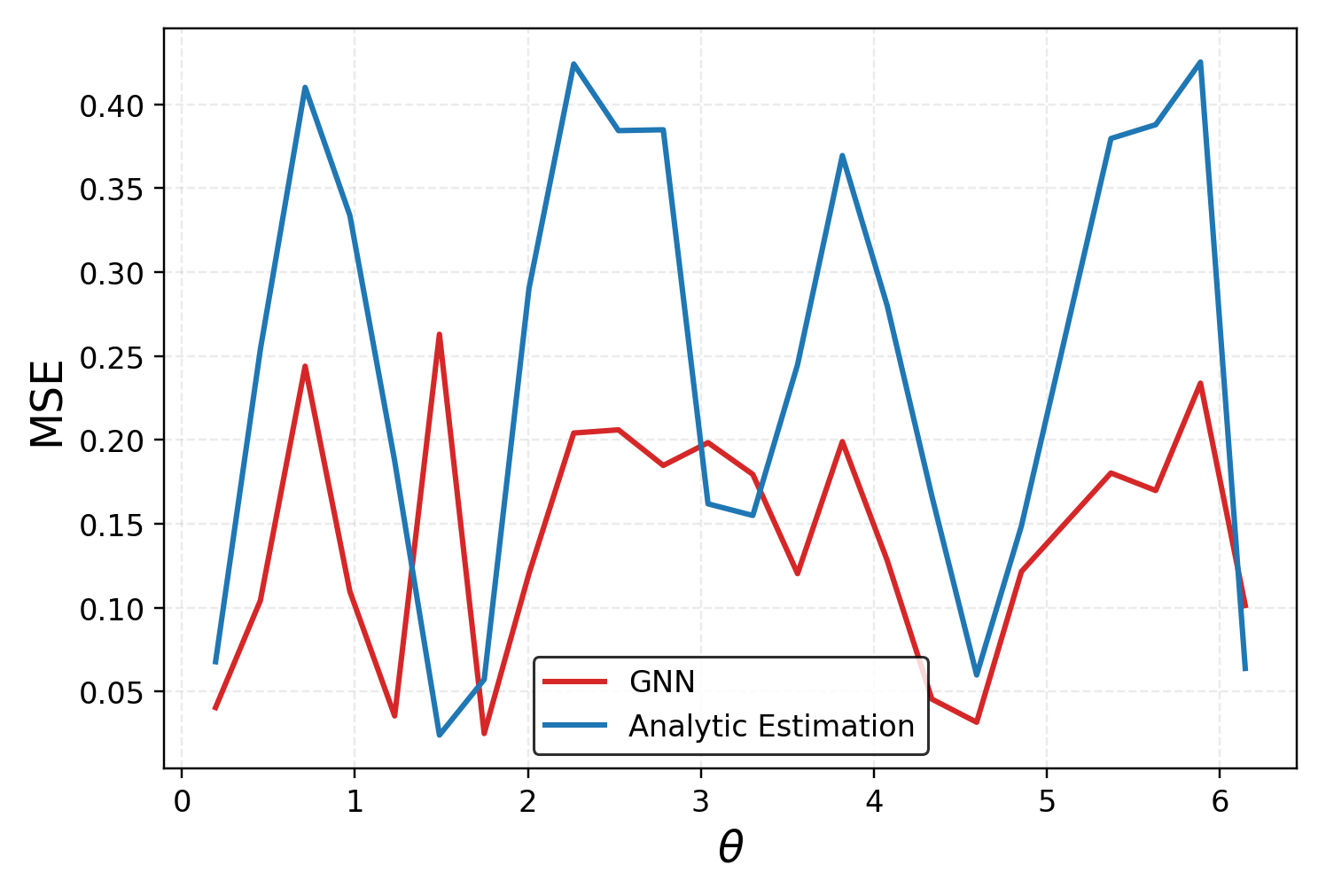}
        \caption{MSE vs $\theta$}
        \label{fig:theta_analytic}
    \end{subfigure}
    \begin{subfigure}[t]{0.49\textwidth}
        \centering
        \includegraphics[width=\linewidth]{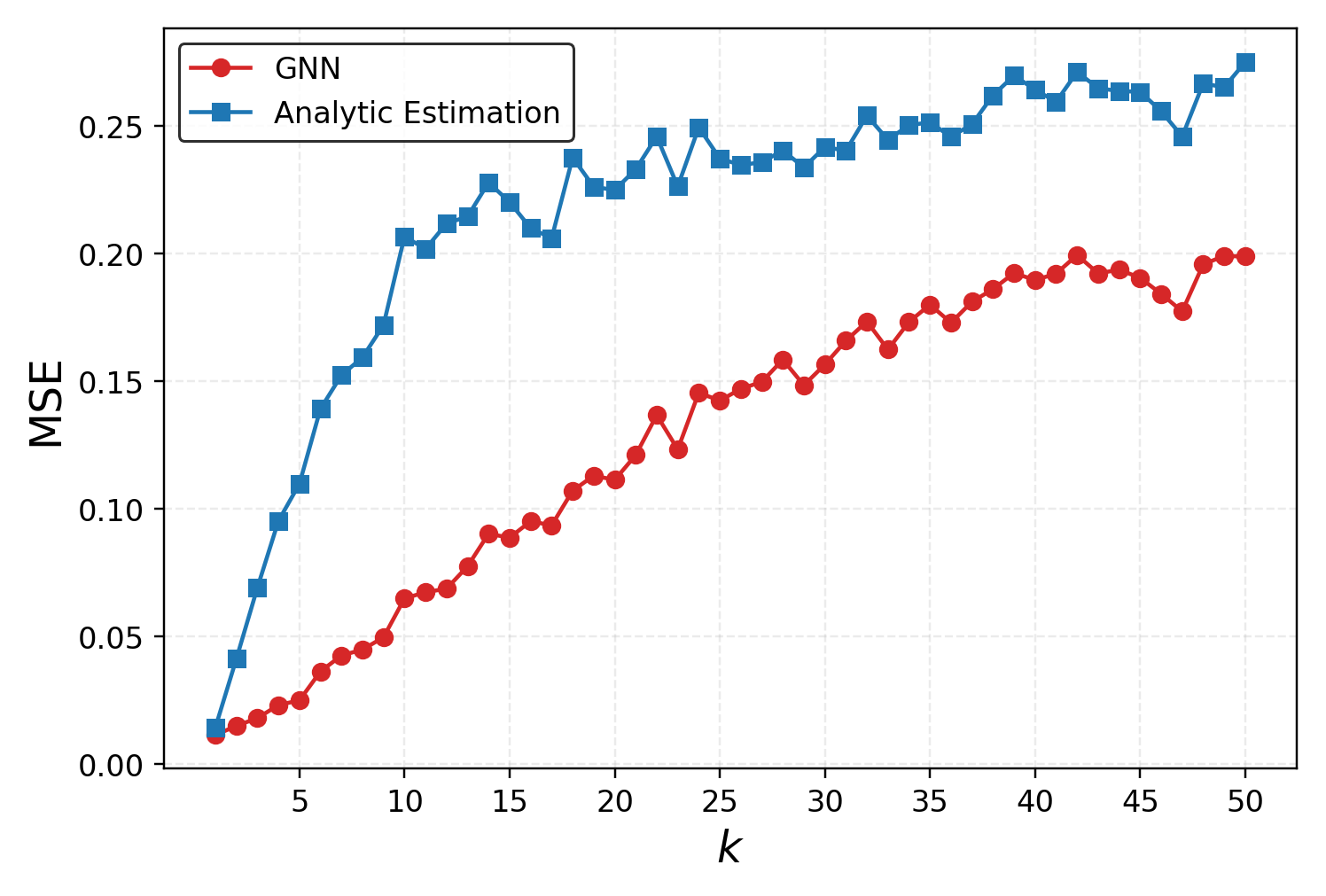}
        \caption{MSE vs $k$}
        \label{fig:k_analytic}
    \end{subfigure}
    \caption{Performance of the GNN model compared to the analytic estimation based on Equation~\ref{eq_13}.}
    \label{fig:results_analytic}
\end{figure}
\end{revision}

\end{document}